\documentclass[5p,twocolumn,times]{elsarticle}
\usepackage{lipsum}
\usepackage{lineno,hyperref}
\modulolinenumbers[5]
\usepackage[pdftex]{color}
\usepackage[font=footnotesize,labelfont=bf]{caption}
\usepackage[font=footnotesize,labelfont=bf]{subcaption}
\journal{Journal of \LaTeX\ Templates}
\usepackage{colortbl}
\usepackage{xcolor}

\usepackage{amssymb}

\biboptions{compress}

\usepackage[figuresright]{rotating}

\begin{document}

\begin{frontmatter}


\title{How multiple weak species jeopardise biodiversity in spatial rock-paper-scissors models}


\address[1]{Institute for Biodiversity and Ecosystem
Dynamics, University of Amsterdam, Science Park 904, 1098 XH
Amsterdam, The Netherlands}
\address[2]{School of Science and Technology, Federal University of Rio Grande do Norte\\
Caixa Postal 1524, 59072-970, Natal, RN, Brazil}

\author[1,2]{J. Menezes}  
\author[2]{R. Barbalho} 

\begin{abstract}
We study generalised rock-paper-scissors models with an arbitrary odd number $N \geq 5$ of species, among which $n$ are weak, with $2 \leq n \leq (N-1)/2$. 
Because of the species' weakness, the probability of individuals conquering territory in the cyclic spatial game is low.
Running stochastic simulations, we study the role of unevenness in the rock-paper-scissors game in spatial patterns and population dynamics, considering diverse models where the weak species 
are in different positions in the cyclic game order. 
Studying systems with with five and seven species, we discover
that the individuals' spatial organisation arising from the pattern formation process determines the stability of the cyclic game with multiple weak species. 
Our outcomes show that the presence of species unbalances the spatial distribution of organisms of the same species bringing consequences on territorial dominance, with the predominant species being determined by the position in the cyclic game order.
Our simulations elucidate that, in general, the further apart the regions inhabited by different weak species are, the less the coexistence between the species is jeopardised. We show that if multiple weak species occupy adjacent spatial domains, the unevenness in the cyclic game is reinforced, maximising the chances of biodiversity loss.
Our discoveries may also be helpful to biologists in comprehending systems where weak species unbalance biodiversity stability.
\end{abstract}


\end{frontmatter}



\section{Introduction}
\label{sec1}

Spatial interactions among species may determine the formation and stability of ecosystems \cite{ecology}. There is plenty of evidence that mobility plays a central role in species coexistence, with animal foraging behaviour, which depends on the environmental conditions, being essential to define conservation strategies \cite{butterfly}. 
The role of space has been reported in many biological systems, which has motivated many authors to give much attention to the cyclic spatial games in ecology \cite{Nature-bio}.
One of the most relevant examples was reported by scientists investigating interactions among three strains of bacteria \textit{Escherichia coli} \cite{bacteria}. First, their experiments revealed that the competition among bacteria is cyclic, thus, being described by the popular rock-paper-scissors game rules. \cite{Coli}.
Second, the authors discovered that the observed cyclic dominance is not sufficient to ensure coexistence, but 
biodiversity is preserved only if individuals interact locally, forming departed spatial domains 
\cite{Allelopathy}. This phenomenon has also been reported 
in competition among groups of lizards and coral reef underwater ecosystems \cite{lizards,Extra1}. This has inspired the formulation of diverse stochastic approaches applying rock-paper-scissors models to simulate spatial interactions and predict the conditions that jeopardise biodiversity or promote species coexistence
\cite{doi:10.1098/rsif.2014.0735,PARK2023113004,PhysRevE.93.062307,doi:10.1063/5.0093342,KABIR2021125767,doi:10.1063/5.0102416,Reichenbach-N-448-1046,Rev1,Bazeia_2017, Avelino-PRE-89-042710,Menezes_2022,Rev6,PhysRevE.99.052310,Rev4,PARKCHAOS,Nagatani2018,Park_2019,RANGEL2022104689,PhysRevE.105.024309,Avelino_2018}.

Extensions of the simplest version of the spatial rock-paper-scissors model have been proposed to investigate more complex systems with a generic number of species \cite{Avelino-PRE-86-031119,2012,Park2017,Pereira,Menezes_2022A}. Local organisms' responses to face epidemic disease outbreaks and nearby enemies in stochastic cyclic models have also been addressed \cite{MENEZES2022104777,TENORIO2022112430,Moura,Anti1,Anti2,MENEZES2022101606}. Furthermore, the rock-paper-scissors game has also been shown to play a fundamental role in the spatial interactions in social systems, public good with punishment, and human bargaining \cite{Rev2,Rev3}.

Because of the relevance of cyclic spatial games to ecosystem stability, the effects of the unbalanced competition capacity affecting organisms of one out of the species have been investigated \cite{uneven,weakest,PedroWeak,Weak4,parity,doi:10.1063/5.0106165}. Studying the simplest rock-paper-scissors model composed of three species, researchers have shown that 
if one out of the species is weaker than the others, this 
species predominates in the spatial game, occupying the most significant fraction of the territory \cite{PedroWeak}. Moreover, if the rock-paper-scissors game unevenness grows, 
the extinction probability accentuates, with the weak species being the most likely to survive \cite{weakest}
This work considers the generalised rock-paper-scissors with an arbitrary odd number of species $N \geq 5$. Our goal is to quantify the impact of multiple weak species on jeopardising biodiversity. 
For this purpose, we assume spatial game systems where the weak species are in different places in the cyclic chain, thus differently unbalancing populating dynamics. 
Running the stochastic simulation, we first quantify the effects of the multiple weak species in systems with five species to understand the pattern formation process. We aim to quantify
the advantages and disadvantages of each species in the competition for space, discovering how the predominant species profit from the unbalanced game with population growth.
Finally, we calculate the coexistence probability considering a wide range of individuals' mobility. Using the results for five species, we anticipate the effects on biodiversity for the general case, where $n$ multiple weak species are present in a system with an odd $N$ number of species, with $2 \leq n \leq (N-1)/2$. We test our prediction for the case with seven species, explaining how the positions of the multiple weak species in the cyclic game
influence coexistence probability, thus, highlighting the case that jeopardises biodiversity the most and the least.

The outline of this paper is as follows. In Sec.~\ref{sec2}, we introduce our methods, detailing 
the implementations of the stochastic simulations. The pattern formation process is addressed in Sec. \ref{sec3}, where the effects of the rock-paper-scissors game unevenness are studied for two models with five species.
Next, the autocorrelation function and the characteristic length size of the typical spatial domains dominated by each species are quantified in Sec.~\ref{sec4}. In Sec.~\ref{sec5}, the 
species' predominance is investigated. We study coexistence probability for systems with five and seven species in Sec.~\ref{sec6}. Our conclusions and discussion appear in
Sec.~\ref{sec7}.

\begin{figure}
    \centering
    \begin{subfigure}{.2\textwidth}
        \centering
        \includegraphics[width=36mm]{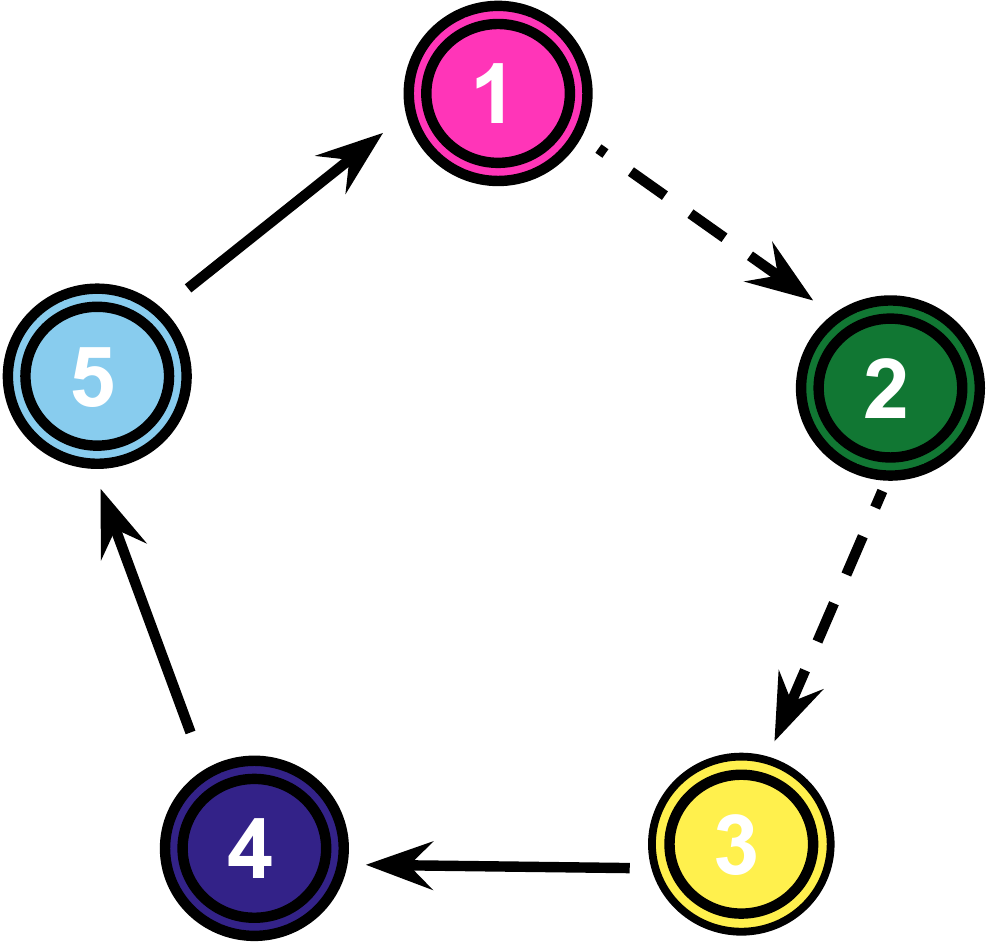}
        \caption{}\label{fig1a}
    \end{subfigure} %
           \begin{subfigure}{.2\textwidth}
        \centering  
        \includegraphics[width=36mm]{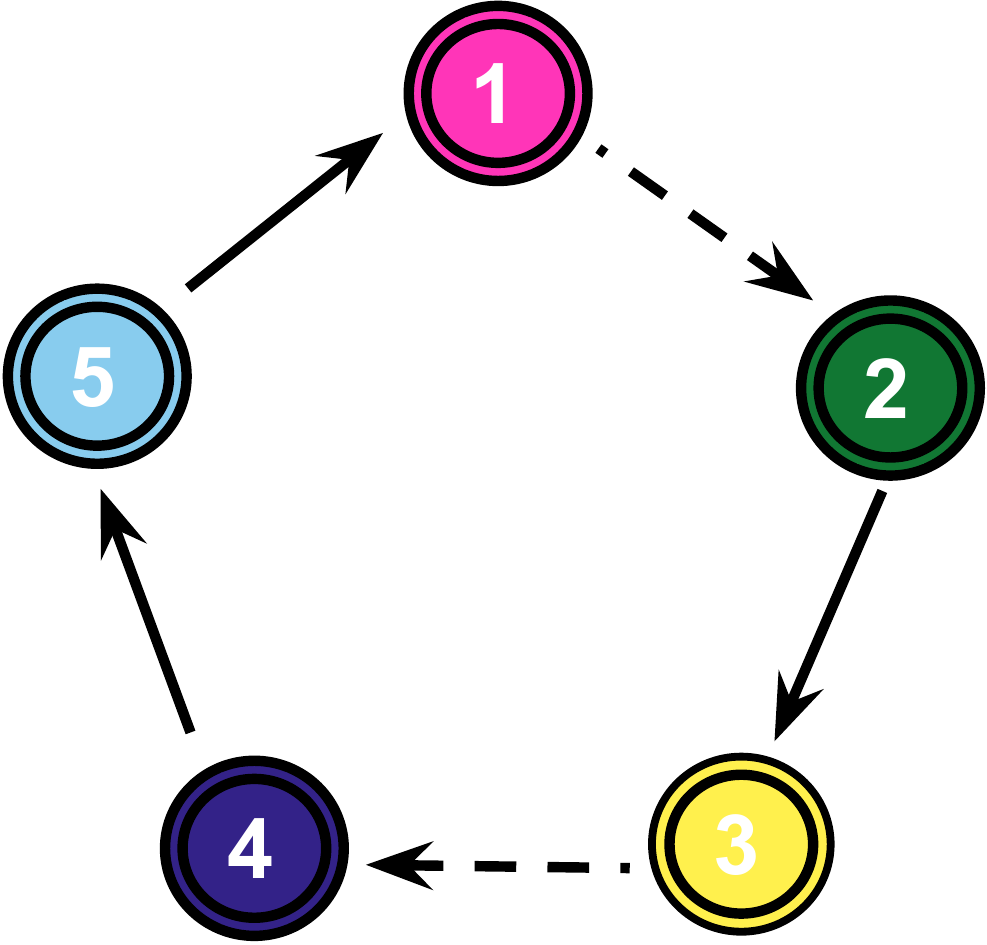}
        \caption{}\label{fig1b}
    \end{subfigure}
\caption{Illustration of the spatial rock-paper-scissors model with five species. 
The colours pink, green, yellow, purple, and light blue represent organisms of species $1$, $2$, $3$, $4$, and $5$, respectively.
The arrows indicate the selection dominance among species, with individuals of species $i$ taking territory of individuals of species $i+1$, for $i=1,2,3,4,5$. 
Figures \ref{fig1a} and \ref{fig1b} show models $5_{\{1,2\}}$ and $5_{\{1,3\}}$, respectively, with dashed lines indicating a reduced selection probability due to the species' weakness.}
\label{fig1}
\end{figure}

\begin{figure*}
	\centering
    \begin{subfigure}{.19\textwidth}
        \centering
        \includegraphics[width=34mm]{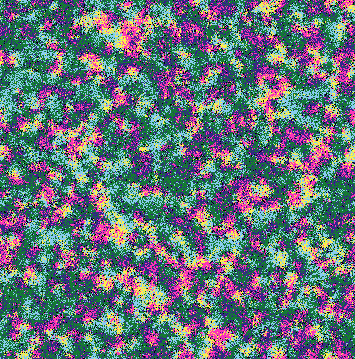}
        \caption{}\label{fig2a}
    \end{subfigure} %
   \begin{subfigure}{.19\textwidth}
        \centering
        \includegraphics[width=34mm]{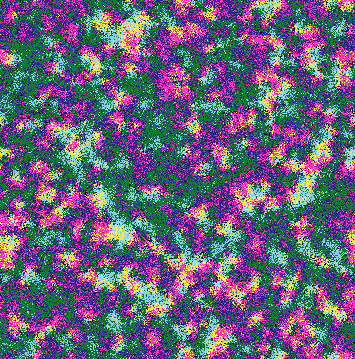}
        \caption{}\label{fig2b}
    \end{subfigure} 
            \begin{subfigure}{.19\textwidth}
        \centering
        \includegraphics[width=34mm]{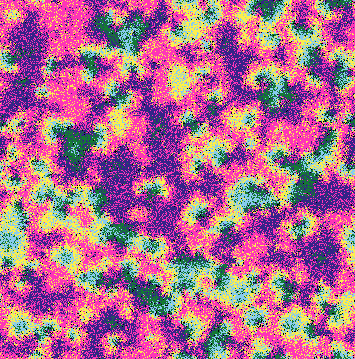}
        \caption{}\label{fig2c}
    \end{subfigure} 
           \begin{subfigure}{.19\textwidth}
        \centering
        \includegraphics[width=34mm]{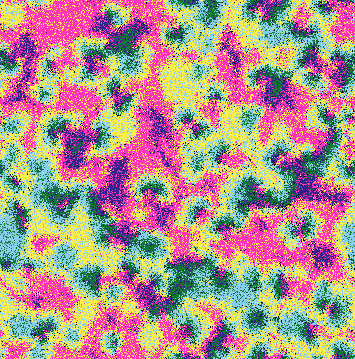}
        \caption{}\label{fig2d}
    \end{subfigure} 
   \begin{subfigure}{.19\textwidth}
        \centering
        \includegraphics[width=34mm]{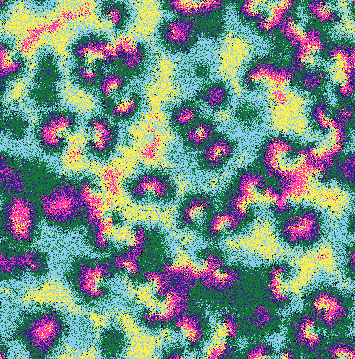}
        \caption{}\label{fig2e}
            \end{subfigure}\\
                \begin{subfigure}{.19\textwidth}
        \centering
        \includegraphics[width=34mm]{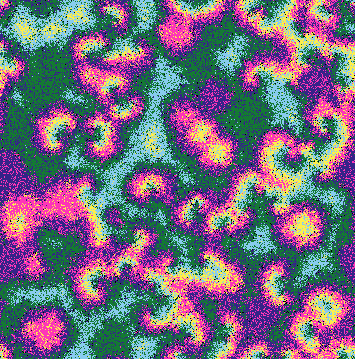}
        \caption{}\label{fig2f}
    \end{subfigure} %
   \begin{subfigure}{.19\textwidth}
        \centering
        \includegraphics[width=34mm]{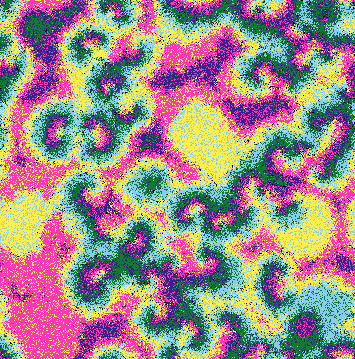}
        \caption{}\label{fig2g}
    \end{subfigure} 
            \begin{subfigure}{.19\textwidth}
        \centering
        \includegraphics[width=34mm]{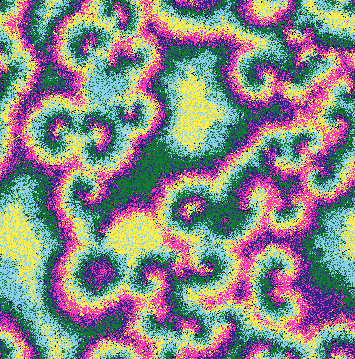}
        \caption{}\label{fig2h}
    \end{subfigure} 
           \begin{subfigure}{.19\textwidth}
        \centering
        \includegraphics[width=34mm]{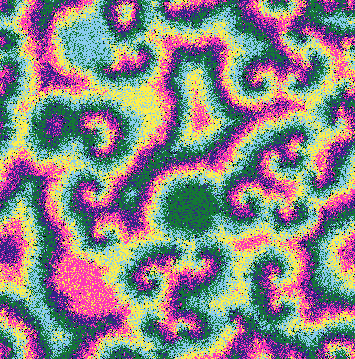}
        \caption{}\label{fig2i}
    \end{subfigure} 
   \begin{subfigure}{.19\textwidth}
        \centering
        \includegraphics[width=34mm]{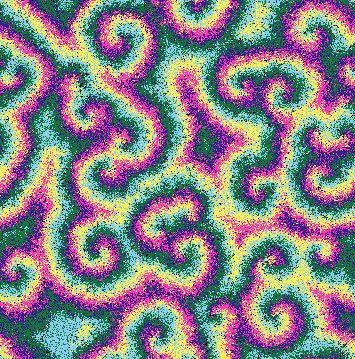}
        \caption{}\label{fig2j}
            \end{subfigure}
 \caption{Snapshots of a simulation starting from random initial conditions of model $5_{\{1,2\}}$, illustrated in Fig.~\ref{fig1a}. The organisms' spatial organisation at $t=60$, $t=120$, $t=180$, 
$t=300$, $t=340$, $t=380$, $t=480$, $t=580$, $t=700$,
and $t=3120$ generations are showed in Figs.~\ref{fig2a}, ~\ref{fig2b},~\ref{fig2c}, ~\ref{fig2d}, ~\ref{fig2e}, ~\ref{fig2f}, ~\ref{fig2g}, ~\ref{fig2h}, ~\ref{fig2i},and ~\ref{fig2j}. 
The realisation was performed in a lattice with $500^2$ grid sites, running for a timespan of $5000$ generations.
The colours follow the scheme in Fig~\ref{fig1}; empty spaces appear as black dots. Video https://youtu.be/ew790sVATAg shows the dynamics of the spatial patterns during the whole simulation.}
  \label{fig2}
\end{figure*}

\section{Methods}
\label{sec2}
In this paper, we study a generalised rock-paper-scissors model with an arbitrary odd number of species $N \geq 5$, where organisms of $n$ species are disadvantaged in the spatial game, with $2 \leq n \leq (N-2)/2$. Let us identify the species using the notation $i$ with $i= 1,...,N$, with the cyclic identification $i=i\,+\,N\,\alpha$ where $\alpha$ is an integer. Our model is denoted by using the notation: $N_{\{weak\,\,\,species\}}$, with the set of weak species being written as an index.

The cyclic models for $N=5$ and $n=2$ are illustrated in Fig.~\ref{fig1}, where arrows indicate the selection dominance, with organisms of species $i$ eliminating individuals of species $i+1$. Figures ~\ref{fig1a} and ~\ref{fig1b} show model $5_{\{1,2\}}$ and $5_{\{1,3\}}$, respectively, with the dashed lines indicating that the probability of organisms of species to compete in the spatial game drops due to an intrinsic species weakness which does not depend on the spatial position \cite{uneven}. 
Although the arrows in Figs.~\ref{fig1a} and \ref{fig1b} indicate that organisms of species $i$ do not attack or are attacked by individuals of species $i\pm2$, they can interact by exchanging positions when moving on the territory. Furthermore, the selection activity of organisms of species $i$ creates empty spaces that can be occupied by individuals of any species.

Our stochastic simulations are performed in
square lattices with periodic boundary conditions, following the May-Leonard numerical implementation, common to studies of spatial games \cite{leonard}. Accordingly,
the total number of individuals is not conserved. Thus, considering that each grid point contains at most one individual; the maximum number of organisms is $\mathcal{N}$, the total number of grid points. We define the density of organisms of species $i$, $\rho_i(t)$, with $i=1,2,3$, as the fraction of the lattice occupied by individuals of the species $i$ at time $t$, $\rho_i(t)=I_i(t)/\mathcal{N}$,  where $I_i(t)$ is the total number of organisms of species $i$ at time $t$. In addition, the temporal dependence of the density of empty spaces is computed as $\rho_0 = 1 - \rho_1 - \rho_2 - \rho_3$.

The initial conditions are prepared by allocating 
one individual at a random grid point. 
We assume that the initial densities of individuals are the same for every species: $\rho_i (t=0) = 1/N$, with $i=1,2,...,N$. For this purpose, throughout this work, all our simulations are performed with the initial number of individuals being the maximum integer number that fits on the lattice, $I_i (t=0)\,\approx \,\mathcal{N}/N$, with $i=1,2,...,N$. The remaining grid sites are left empty in the initial conditions. However, we have repeated 
our simulations for various initial proportions of empty sites. We concluded that our main conclusions do not depend on the density of vacancies in the initial conditions.

Once the random initial conditions are built, the spatial interactions are stochastically implemented as follows:
\begin{itemize}
\item 
Selection: $ i\ j \to i\ \otimes\,$, with $ j = i+1$, where $\otimes$ means an empty space. This means that during a selection interaction, which obeys the generalised rock-paper-scissors game rules,
the organism of species $i+1$ disappears, leaving an empty space;
\item
Reproduction: $ i\ \otimes \to i\ i\,$. When a reproduction interaction occurs, a new organism of species $i$ occupies the available empty space.
\item 
Mobility: $ i\ \odot \to \odot\ i\,$, where $\odot$ means either an organism of any species. For a mobility interaction, an individual of species $i$ exchanges positions with either another organism of any species or an empty space.
\end{itemize}

As we implement the Moore neighbourhood, individuals may interact with one of their eight nearest neighbours.
The probability of an interaction being raffled in the stochastic process is $s$, $r$ and $m$, for selection, reproduction, and mobility, respectively. Although the interaction probability is the same for every species, the implementation depends on the species' strength whenever a selection interaction is randomly chosen. Because of this, we introduce the strength factor $\kappa_i$, with $0 \leq \kappa_i \leq 1$, a real parameter to control the selection interaction implementation for species organisms $i$. Throughout this paper, we consider that for all organisms of the multiple weak species, the strength factor is given by $\kappa$, independent of the species - for strong species, the strength factor is maximum \cite{uneven}.

For implementing the spatial interactions, the algorithm randomly chooses one individual among all organisms of every species to execute one of the interactions, which is raffled according to the set of probabilities. The individual that suffers the interaction is one of the eight immediate neighbours, which is randomly chosen.
Furthermore, every time a selection interaction is drawn, the probability of execution depends on the species strength factor $\kappa_i$. 
If the interaction is implemented, one timestep is counted. Otherwise, the steps are repeated. Our time unit is named generation, which is the time spent to $\mathcal{N}$ timesteps to occur.




\section{Spatial patterns}
\label{sec3}
Let us first investigate the impact of multiple weak species on the pattern formation process. For this purpose, we consider a single realisation starting from random initial conditions in 
a square lattice with $500^2$ grid points for models $5_{\{1,2\}}$ and $5_{\{1,3\}}$, illustrated in Figs.~\ref{fig1a} and \ref{fig1b}, respectively. The timespan is $5000$ generations; the strength factor for organisms of the multiple weak species is given by $\kappa=0.75$. To improve the visualisation of the organisms' spatial organisation, we used the interaction probabilities: $s=r=0.35$ and $m=0.3$; however, we have performed simulations with other sets of probabilities and verified that our conclusions are independent of the model parameters.

\subsection{Model $5_{\{1,2\}}$}

Figures \ref{fig2a}, \ref{fig2b}, \ref{fig2c}, \ref{fig2d}, \ref{fig2e}, \ref{fig2f}, \ref{fig2g}, \ref{fig2h}, \ref{fig2i}, and \ref{fig2j} show snapshots of a realisation of model $5_{\{1,2\}}$, captured after $60$, $120$, $180$, $300$, $340$, $380$, $480$, $580$, $700$, and $3120$ generations, respectively. The dynamics 
of the organisms' spatial organisation during the entire simulation are shown in the video https://youtu.be/ew790sVATAg. 

Figure \ref{fig2a} shows that groups of individuals of two species are formed as soon as the simulation commences: $(1,3)$, $(4,1)$, $(2,4)$, $(5,2)$, and $(3,5)$. This happens because the organisms of species $i$ and $i+2$ do not attack each other, thus peacefully sharing local spatial domains.
Furthermore, we observe that the abundances of the species sharing a patch are not the same but higher for species $i$ than for species $i+2$. The spatial domains with two non-interacting species grow, with an alternate local dominance obeying the following order $\{(1,3);(3,5),(2,5),(2,4),(1,4)\}$,
as shown in Figs. \ref{fig2b} to \ref{fig2d}. 

Subsequently, the local alternate growth of two-species spatial domains is interrupted when stochastic fluctuations form spirals in various parts of the lattices, allowing species to coexist, as observed in Fig.~\ref{fig2e}. The spirals then grow and spread on the lattice, as shown in the snapshots in Figs. \ref{fig2f} to \ref{fig2j}; the spiral arms are formed by organisms of species that do not interact in the spatial game, with wavefronts in the following order: $\{(1,3);(3,5),(2,5),(2,4),(1,4)\}$ \cite{2012,Moura}. 
\begin{figure*}
	\centering
    \begin{subfigure}{.19\textwidth}
        \centering
        \includegraphics[width=34mm]{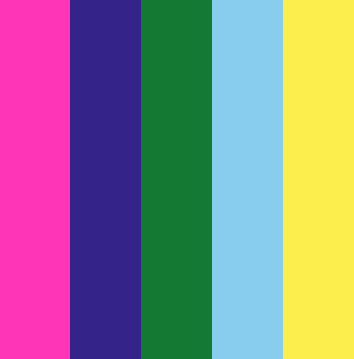}
        \caption{}\label{fig3a}
    \end{subfigure} %
   \begin{subfigure}{.19\textwidth}
        \centering
        \includegraphics[width=34mm]{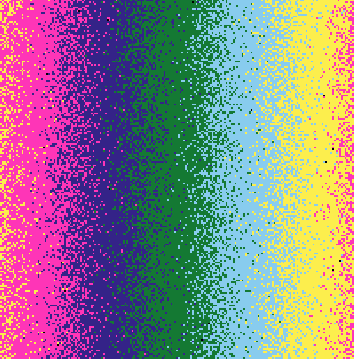}
        \caption{}\label{fig3b}
    \end{subfigure} 
            \begin{subfigure}{.19\textwidth}
        \centering
        \includegraphics[width=34mm]{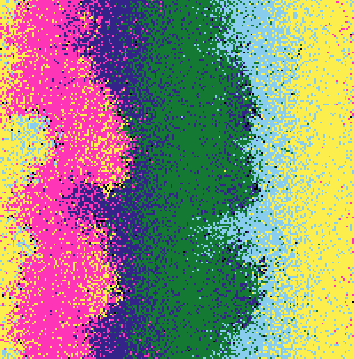}
        \caption{}\label{fig3c}
    \end{subfigure} 
           \begin{subfigure}{.19\textwidth}
        \centering
        \includegraphics[width=34mm]{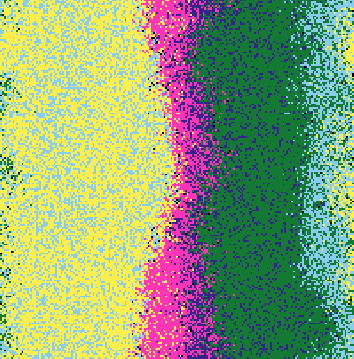}
        \caption{}\label{fig3d}
    \end{subfigure} 
   \begin{subfigure}{.19\textwidth}
        \centering
        \includegraphics[width=34mm]{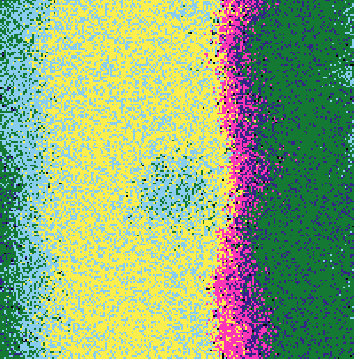}
        \caption{}\label{fig3e}
            \end{subfigure}\\
                \begin{subfigure}{.19\textwidth}
        \centering
        \includegraphics[width=34mm]{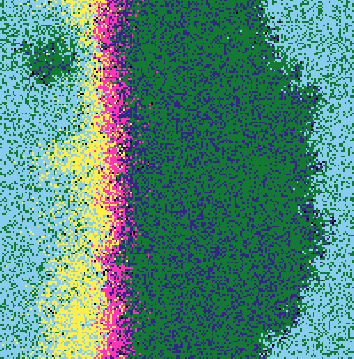}
        \caption{}\label{fig3f}
    \end{subfigure} %
   \begin{subfigure}{.19\textwidth}
        \centering
        \includegraphics[width=34mm]{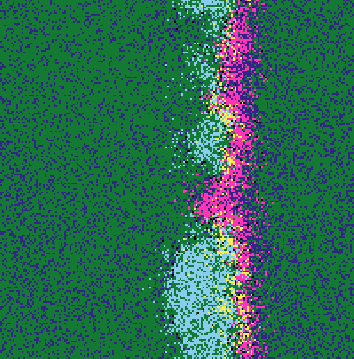}
        \caption{}\label{fig3g}
    \end{subfigure} 
            \begin{subfigure}{.19\textwidth}
        \centering
        \includegraphics[width=34mm]{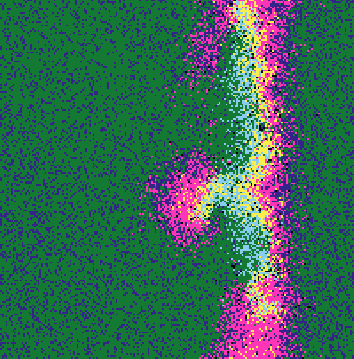}
        \caption{}\label{fig3h}
    \end{subfigure} 
           \begin{subfigure}{.19\textwidth}
        \centering
        \includegraphics[width=34mm]{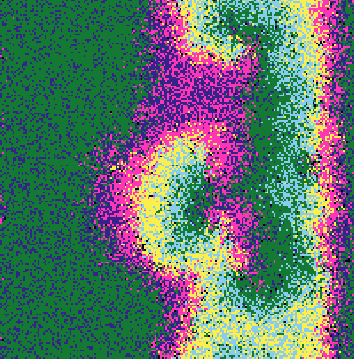}
        \caption{}\label{fig3i}
    \end{subfigure} 
   \begin{subfigure}{.19\textwidth}
        \centering
        \includegraphics[width=34mm]{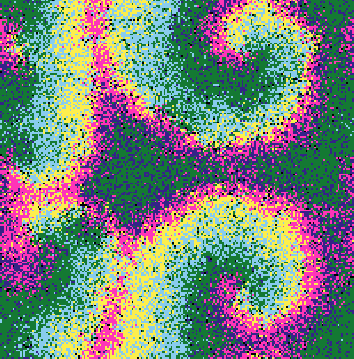}
        \caption{}\label{fig3j}
            \end{subfigure}
 \caption{Snapshots of a simulation of model $5_{\{1,2\}}$ starting from 
the prepared initial conditions in Fig.~\ref{fig3a} in a lattice with $200^2$ grid sites for a timespan of $3500$ generations.
The organisms' spatial organisation at $t=154$, $t=266$, $t=462$, 
$t=588$, $t=1134$, $t=1470$, $t=1568$, $t=1708$,
and $t=2828$ generations are showed in Figs.~\ref{fig3b}, ~\ref{fig3c}, ~\ref{fig3d}, ~\ref{fig3e}, ~\ref{fig3f}, ~\ref{fig3g}, ~\ref{fig3h}, ~\ref{fig3i},and ~\ref{fig3j}. The colours follow the scheme in Fig~\ref{fig1}; empty spaces appear as black dots. Video https://youtu.be/IQkPxMO63nw shows the dynamics of the spatial patterns during the whole simulation.}
  \label{fig3}
\end{figure*}

To understand the process that interrupts the alternate local dominance, producing spiral waves in the rock-paper-scissors game with multiple weak species, we ran simulations starting from the configuration in Fig.~\ref{fig3a}, with organisms of a single species filling torus ring surfaces with the same width. The sequence of colours shows that initial conditions were purposely prepared so that non-interacting species form that adjacent rings, namely, species $1$ (pink), $4$ (purple), $2$ (green), $5$ (light blue), and $3$ (yellow). The realisations ran in lattices with $200^2$ grid sites for a timespan of $3500$ generations; the strength factor of the multiple weak species is $\kappa=0.25$.

As soon as the simulation starts, individuals disperse through adjacent rings, sharing space without aggression, as shown in Fig.~\ref{fig3b}. When individuals of species $i$ meet individuals of species $i+1$, selection interactions occur, as observed in Fig.~\ref{fig3c}, where black dots show empty spaces.
This provokes the torus surface rotation from left to right, as observed in the video https://youtu.be/IQkPxMO63nw. However, the unevenness in the cyclic game produces a delay in advance of rings mostly occupied by weak species \cite{uneven}. For this reason, the average width of the area with species $1$ narrows, as shown in Fig.~\ref{fig3d}. As the invasion rate of species $1$ over species $2$ is low, the area with individuals of species $2$ wides, even though species $2$ is also weak. 
The outcomes show that organisms of species $2$ stochastically manage to move without being caught by individuals of species $1$, reaching the area dominated by species $3$ and $5$. 
At this point, species $2$ proliferates (green) by eliminating individuals of species $3$ (yellow), as shown in Fig.~\ref{fig3e}. Subsequently, organisms of species $4$ (purple) also cross the barrier of species $3$ (yellow), thus invading the area of species $2$ and $5$. The result is that the torus becomes mostly dominated by species $2$ and $4$, with four narrow patches of species $(2,5)$, $(3,5)$, $(1,3)$ and $(1.4)$, as shown in Figs.~\ref{fig3f} and \ref{fig3g}. This spatial configuration facilitates that stochastic oscillations lead individuals of species $1$ (pink) to reach organisms of species $2$ (green), being followed by individuals of the other species, thus creating a spiral wave that spreads on the lattice, as appears in Figs. \ref{fig3h} to \ref{fig3j}. Once spirals arise, species' strength factors determine population dynamics.

\subsection{Model $5_{\{1,3\}}$}
\begin{figure*}
	\centering
    \begin{subfigure}{.19\textwidth}
        \centering
        \includegraphics[width=34mm]{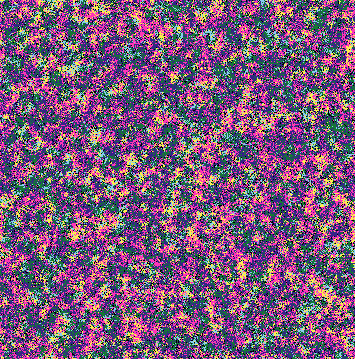}
        \caption{}\label{fig4a}
    \end{subfigure} %
   \begin{subfigure}{.19\textwidth}
        \centering
        \includegraphics[width=34mm]{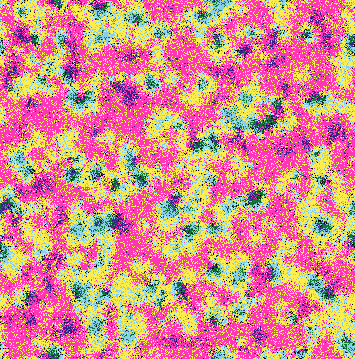}
        \caption{}\label{fig4b}
    \end{subfigure} 
            \begin{subfigure}{.19\textwidth}
        \centering
        \includegraphics[width=34mm]{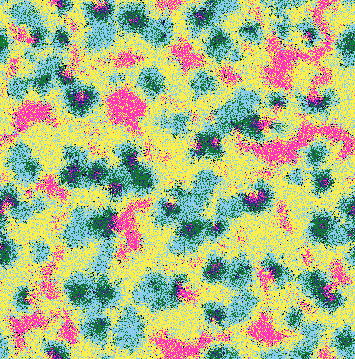}
        \caption{}\label{fig4c}
    \end{subfigure} 
           \begin{subfigure}{.19\textwidth}
        \centering
        \includegraphics[width=34mm]{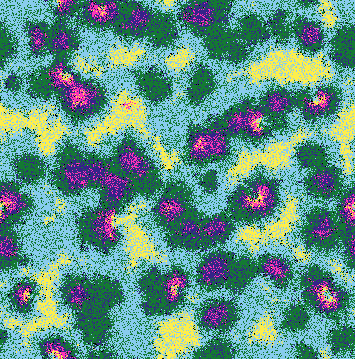}
        \caption{}\label{fig4d}
    \end{subfigure} 
   \begin{subfigure}{.19\textwidth}
        \centering
        \includegraphics[width=34mm]{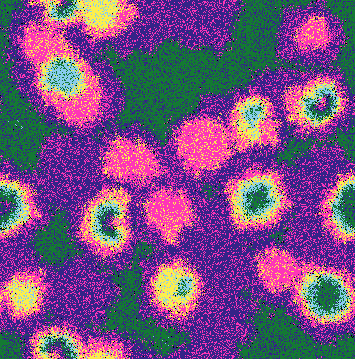}
        \caption{}\label{fig4e}
            \end{subfigure}\\
                \begin{subfigure}{.19\textwidth}
        \centering
        \includegraphics[width=34mm]{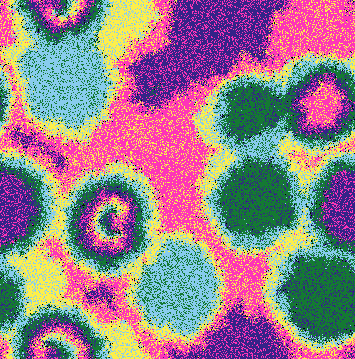}
        \caption{}\label{fig4f}
    \end{subfigure} %
   \begin{subfigure}{.19\textwidth}
        \centering
        \includegraphics[width=34mm]{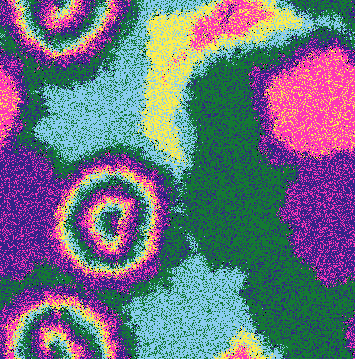}
        \caption{}\label{fig4g}
    \end{subfigure} 
            \begin{subfigure}{.19\textwidth}
        \centering
        \includegraphics[width=34mm]{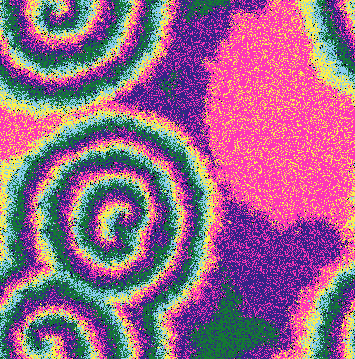}
        \caption{}\label{fig4h}
    \end{subfigure} 
           \begin{subfigure}{.19\textwidth}
        \centering
        \includegraphics[width=34mm]{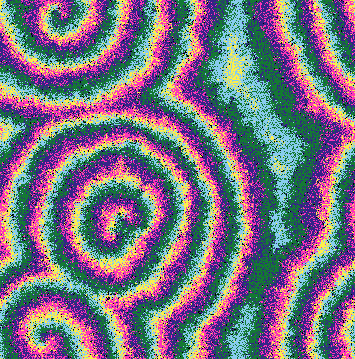}
        \caption{}\label{fig4i}
    \end{subfigure} 
   \begin{subfigure}{.19\textwidth}
        \centering
        \includegraphics[width=34mm]{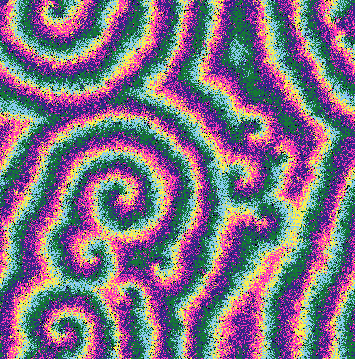}
        \caption{}\label{fig4j}
            \end{subfigure}
 \caption{Snapshots of a simulation starting from random initial conditions of model $5_{\{1,3\}}$, illustrated in Fig.~\ref{fig1b}. 
The realisation ran in a lattice with $500^2$ grid sites for a timespan of $5000$ generations.
Figures ~\ref{fig4a}, ~\ref{fig4b},~\ref{fig4c}, ~\ref{fig4d}, ~\ref{fig4e}, ~\ref{fig4f}, ~\ref{fig4g}, ~\ref{fig4h}, ~\ref{fig4i},and ~\ref{fig4j} depict the individuals' spatial organisation at $t=20$, $t=60$, $t=100$, $t=160$, $t=300$, $t=460$, $t=640$, $t=960$, $t=1320$, and $t=3380$ generations. Video https://youtu.be/CXhha2HhNek shows the changes in the spatial configuration during the simulation. The colours follow the scheme in Fig~\ref{fig1}; black dots show the empty sites. }
  \label{fig4}
\end{figure*}

Figures \ref{fig4a}, \ref{fig4b}, \ref{fig4c}, \ref{fig4d}, \ref{fig4e}, \ref{fig4f}, \ref{fig4g}, \ref{fig4h}, \ref{fig4i}, and \ref{fig4j} depict the spatial organisation observed in the simulation of model $5_{\{1,3\}}$, captured after $20$, $60$, $100$, $160$, $300$, $460$, $640$, $960$, $1320$, and $3380$ generations, respectively. The dynamics 
of the organisms' spatial organisation during the entire simulation are shown in the video https://youtu.be/CXhha2HhNek.
According to Figs.~\ref{fig4a} to \ref{fig4d}, the weakness of species $1$ and $3$ generates an alternate territory dominance in the initial simulation stage, which is similar to the phenomenon observed in the pattern formation period of model $5_{\{1,2\}}$.
After that, spiral waves arise, as observed in Figs.~\ref{fig4e} to \ref{fig4j}.



\begin{figure}[h]
 \centering
        \begin{subfigure}{.48\textwidth}
        \centering
        \includegraphics[width=90mm]{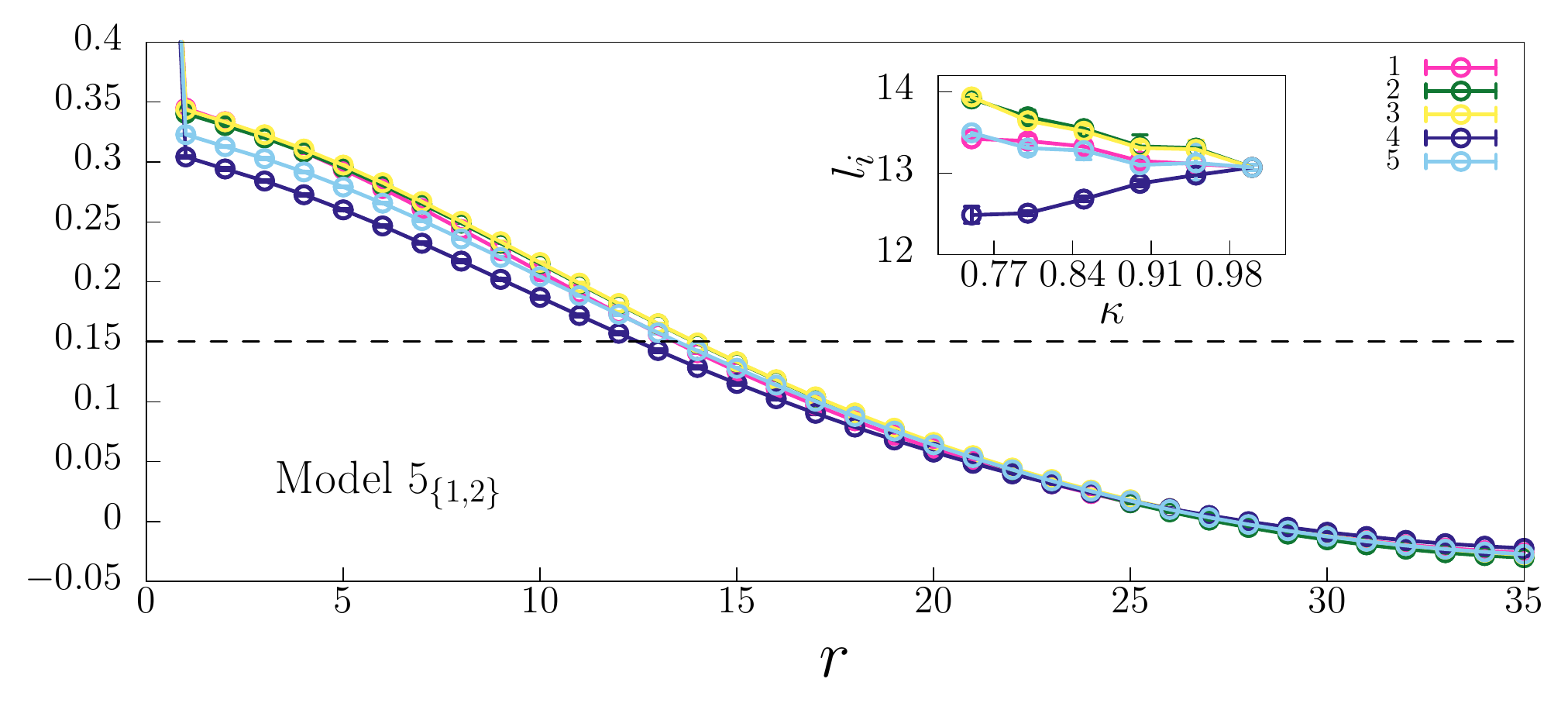}
        \caption{}\label{fig5a}
    \end{subfigure}\\
       \begin{subfigure}{.48\textwidth}
        \centering
        \includegraphics[width=90mm]{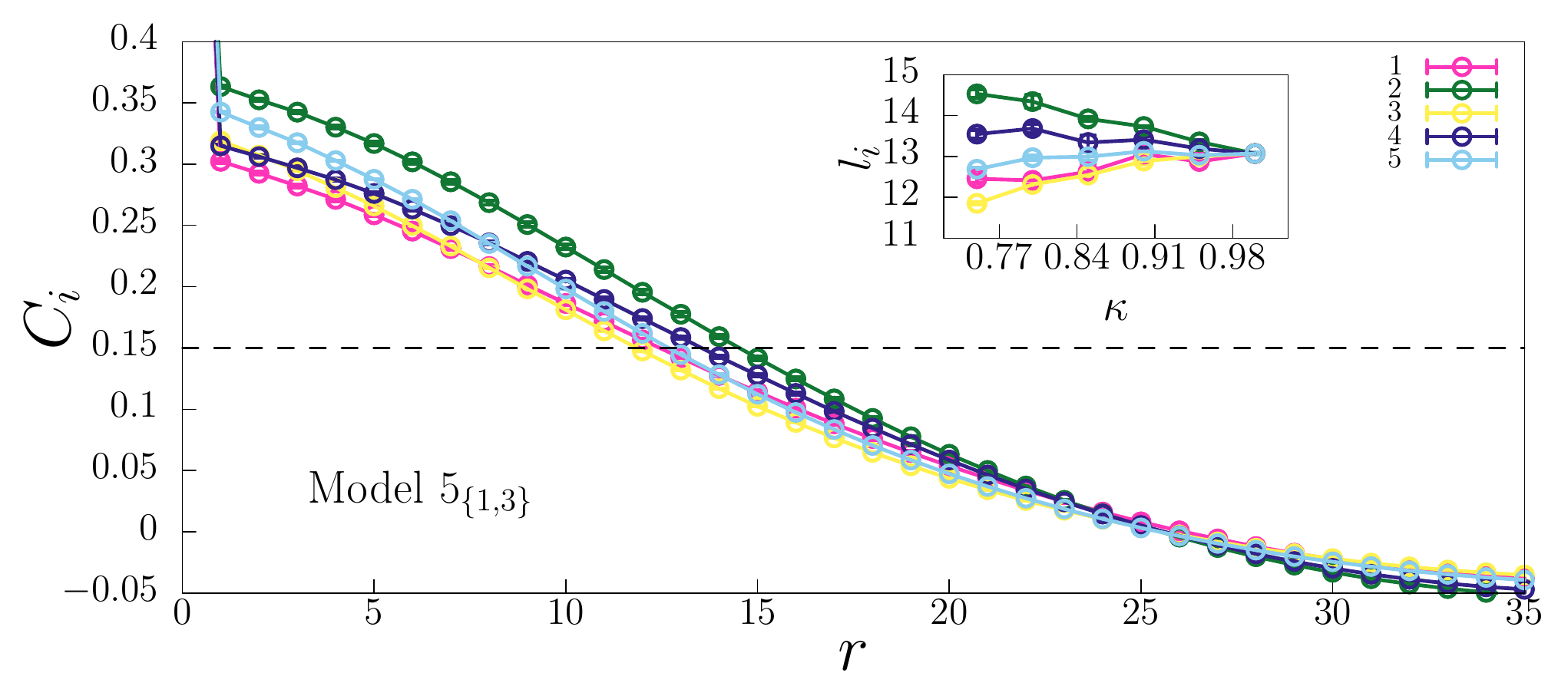}
        \caption{}\label{fig5b}
    \end{subfigure}
\caption{Autocorrelation functions $C_i(r)$ and characteristic length $l_i$ in terms of the weak species' strength factor. Figures \ref{fig5a} and \ref{fig5b} show the 
mean autocorrelation function for $\kappa=0.75$ averaged from a set of $100$ simulations, for models $5_{\{1,2\}}$ and $5_{\{1,3\}}$, respectively.
The inset figure 
shows how the characteristic length scales depends on $\kappa$. The error bars indicate the standard deviation; 
colours follow the scheme in Fig.~\ref{fig1}. The horizontal dashed black line indicates the threshold assumed to calculate the characteristic length.}
  \label{fig5}
\end{figure}

\begin{figure}
    \centering
        \begin{subfigure}{.48\textwidth}
        \centering
        \includegraphics[width=85mm]{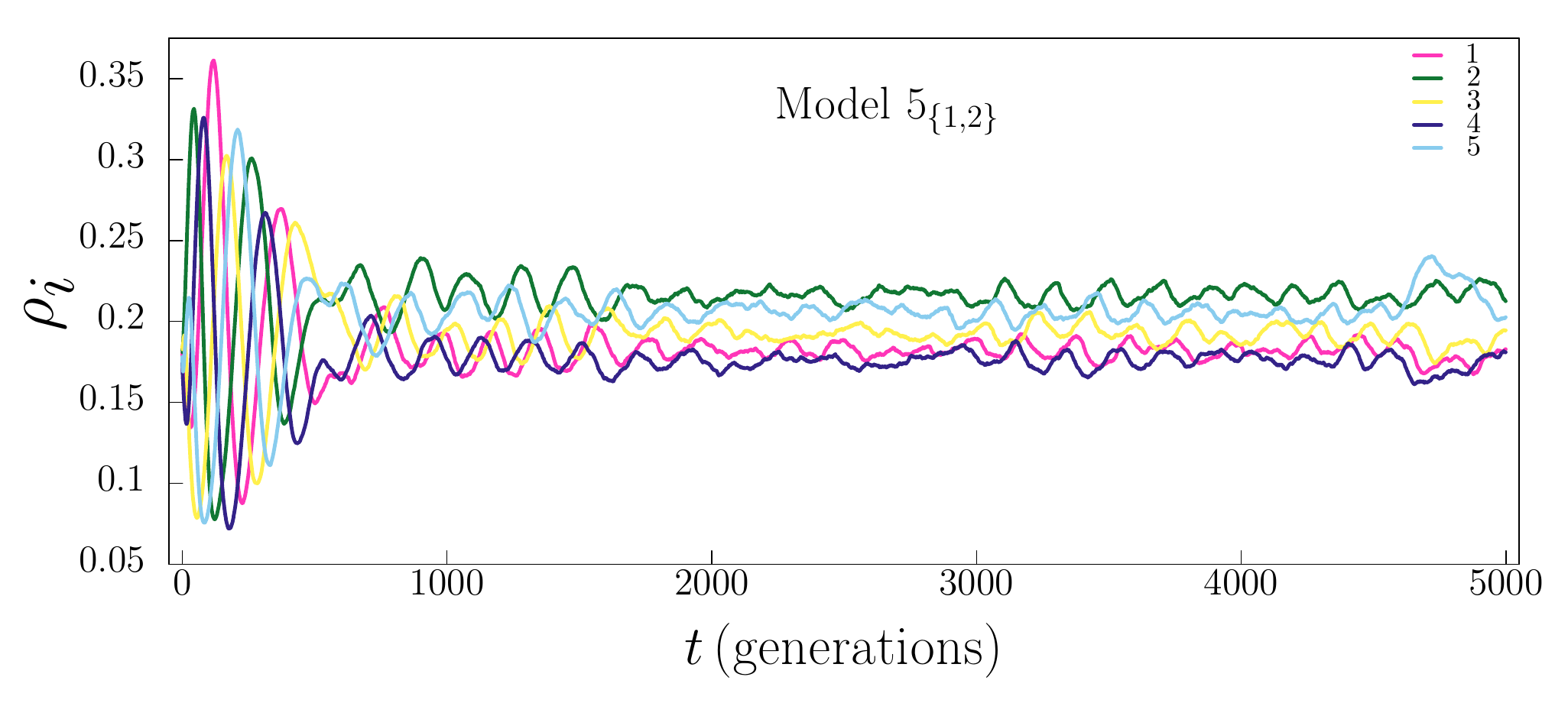}
        \caption{}\label{fig6a}
    \end{subfigure}\\
       \begin{subfigure}{.48\textwidth}
        \centering
        \includegraphics[width=85mm]{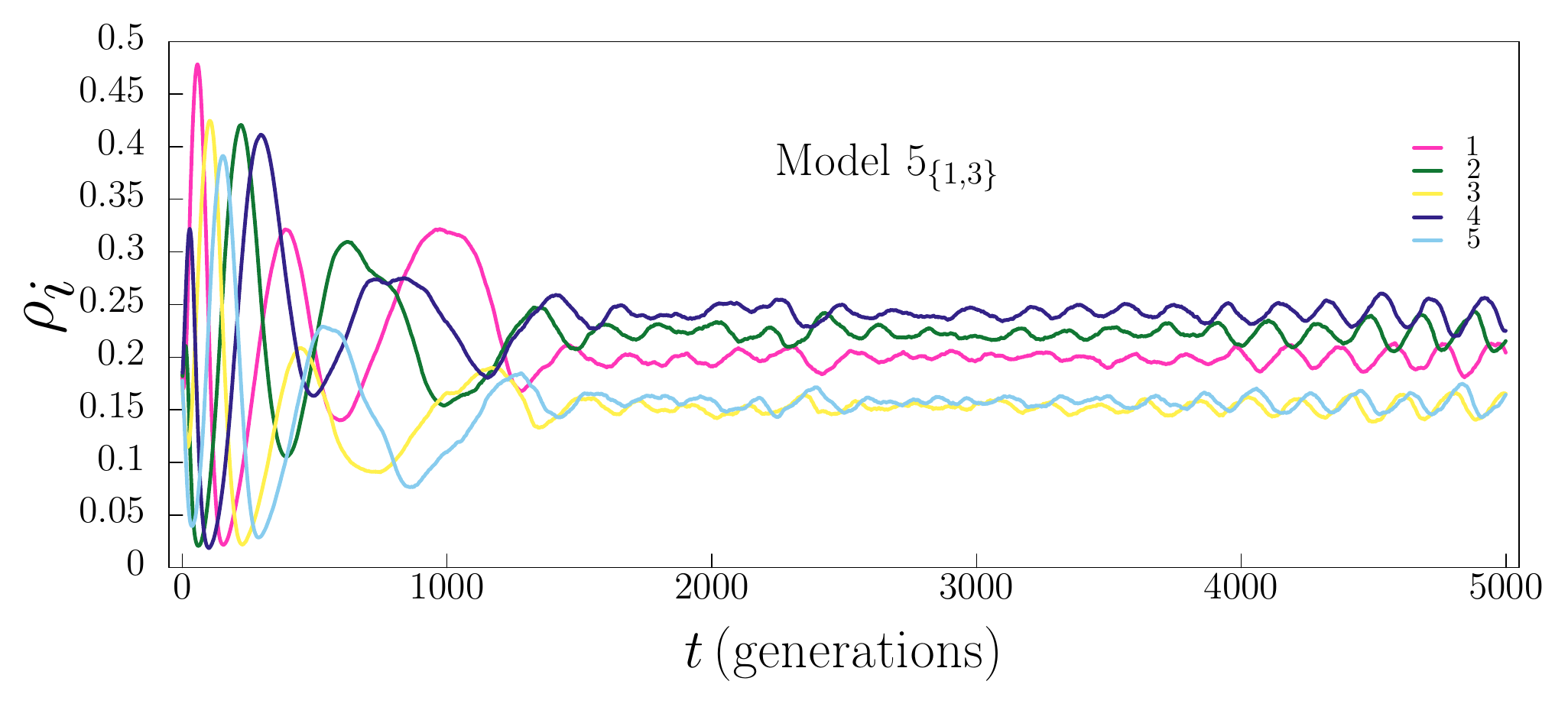}
        \caption{}\label{fig6b}
    \end{subfigure}
\caption{Dynamics of the species densities during the course of the simulations presented in Figs.~\ref{fig3} (https://youtu.be/ew790sVATAg) and ~\ref{fig5} (https://youtu.be/CXhha2HhNek).
The colours follow the scheme in Fig.~\ref{fig1}.
}
\label{fig6}
\end{figure}

\begin{figure}[t]
 \centering
       \begin{subfigure}{.48\textwidth}
        \centering
        \includegraphics[width=85mm]{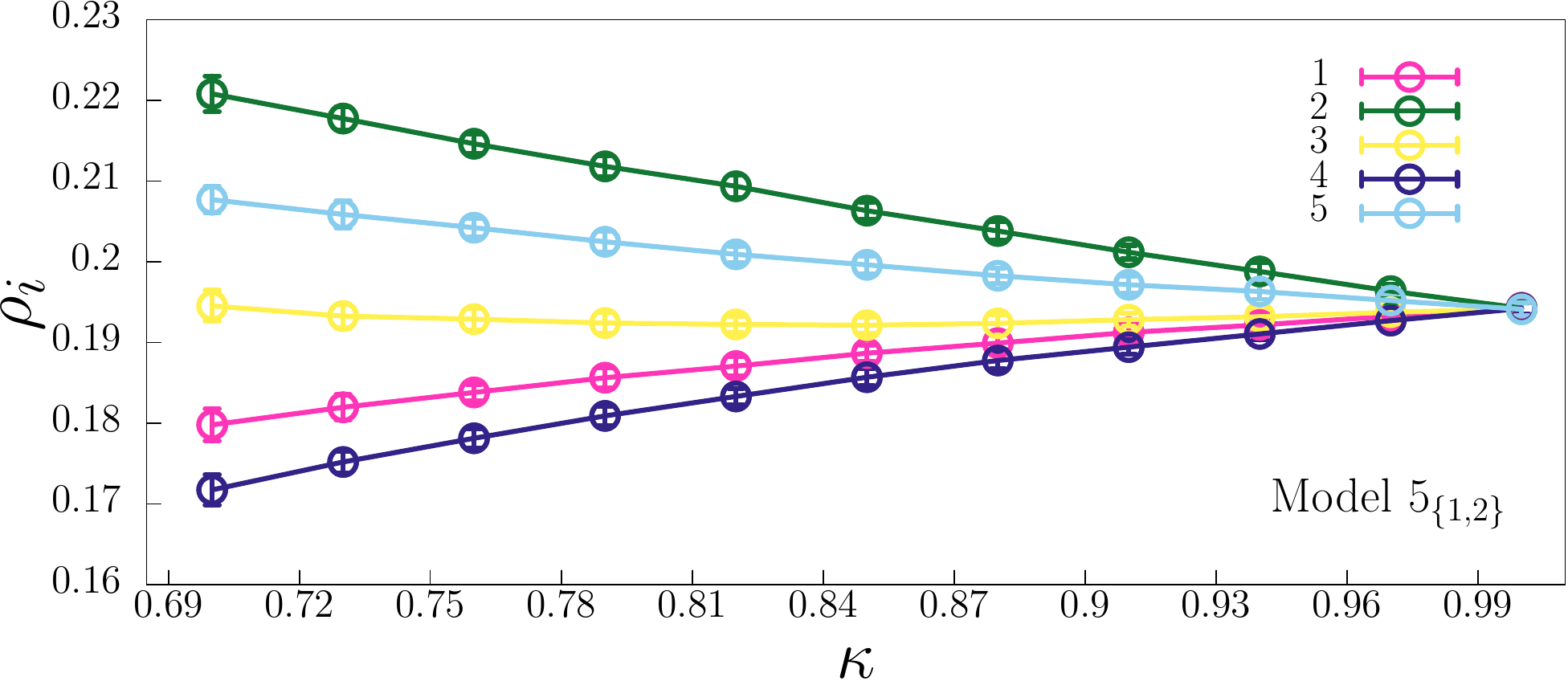}
        \caption{}\label{fig7a}
    \end{subfigure}\\
           \begin{subfigure}{.48\textwidth}
        \centering
        \includegraphics[width=85mm]{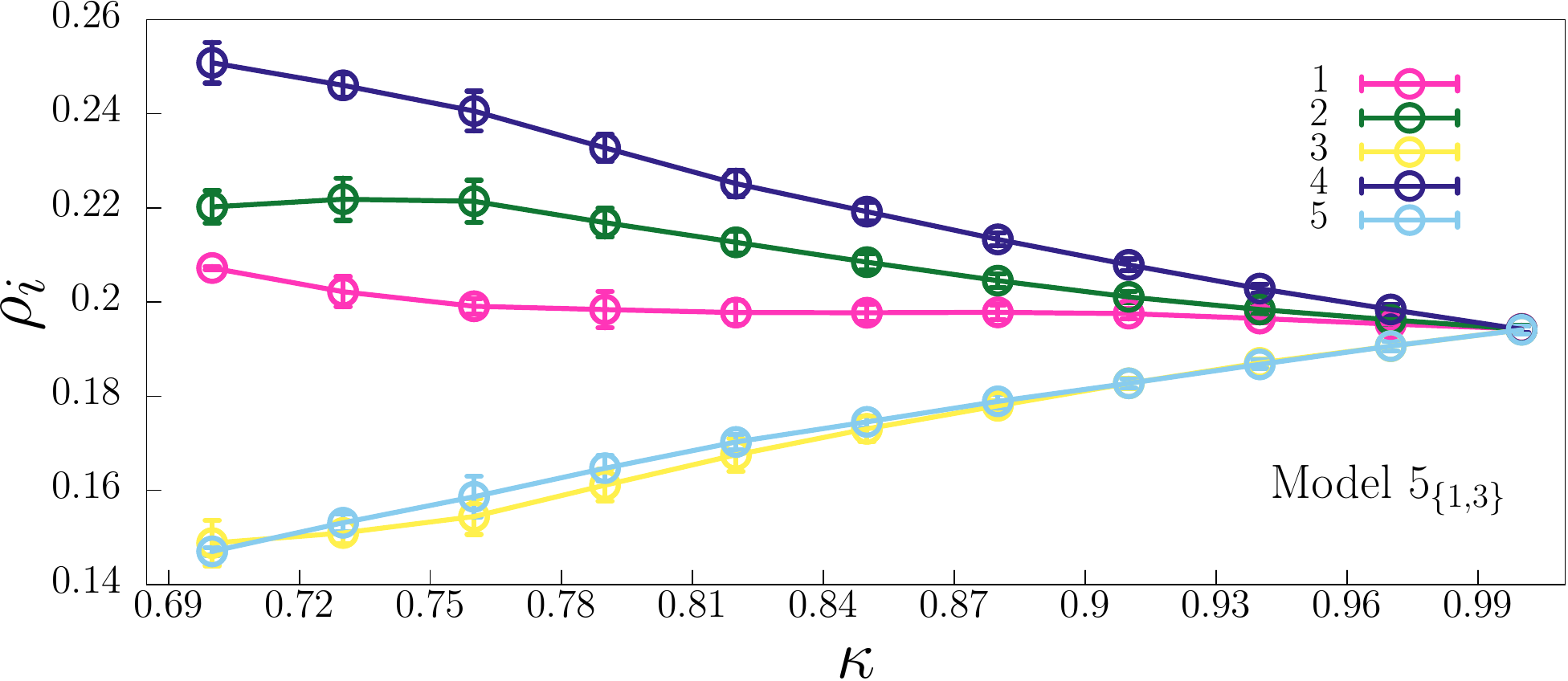}
        \caption{}\label{fig7b}
    \end{subfigure}
\caption{Species densities as functions of the strength factor.
Figures~\ref{fig7a} and \ref{fig7b} show the results for models $5_{\{1,2\}}$ and $5_{\{1,3\}}$.
The outcomes were averaged from sets of $100$ simulations, starting from different initial conditions;
the error bars show the standard deviation. The colours follow the scheme in Fig.~\ref{fig1}.}
  \label{fig7}
\end{figure}

\begin{figure}[h]
 \centering
        \begin{subfigure}{.48\textwidth}
        \centering
        \includegraphics[width=85mm]{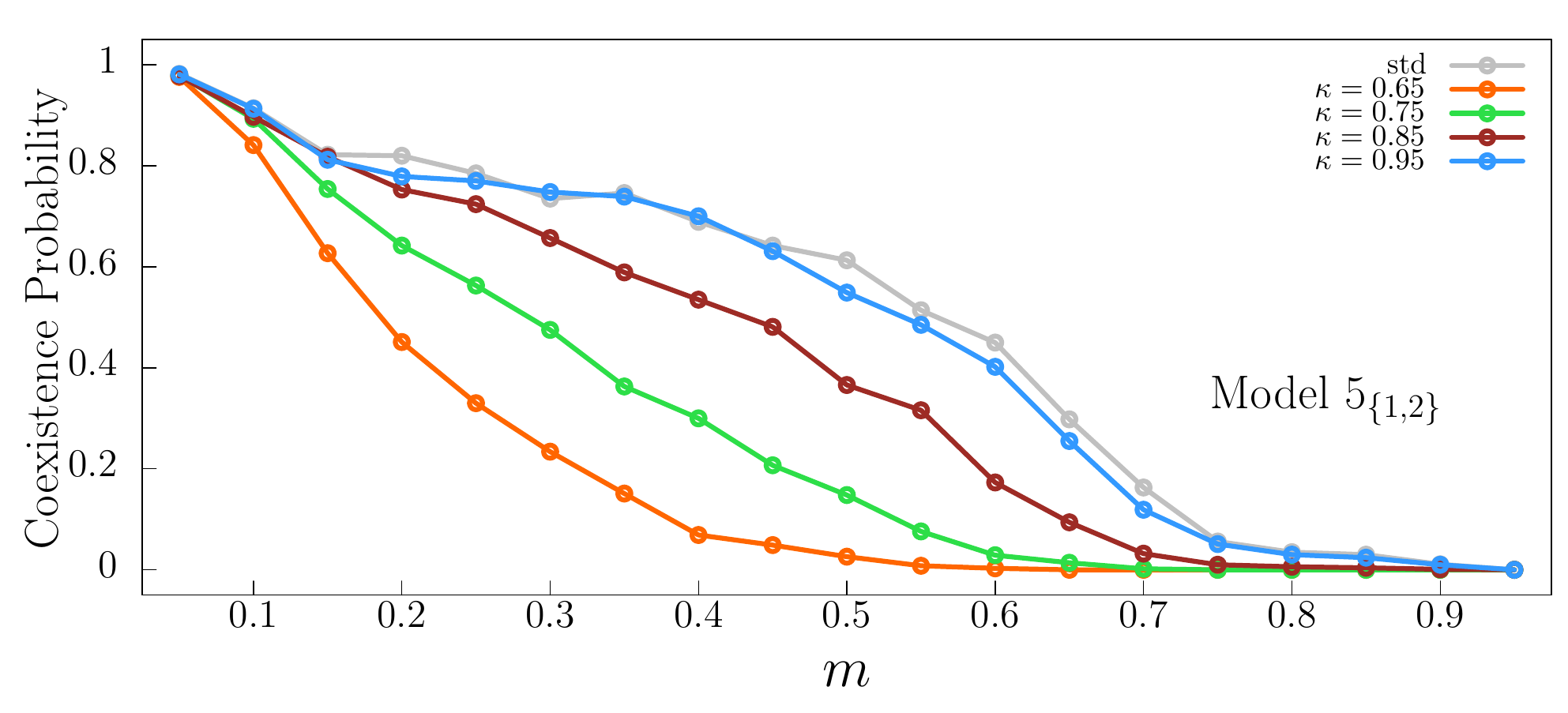}
        \caption{}\label{fig8a}
    \end{subfigure}\\
       \begin{subfigure}{.48\textwidth}
        \centering
        \includegraphics[width=85mm]{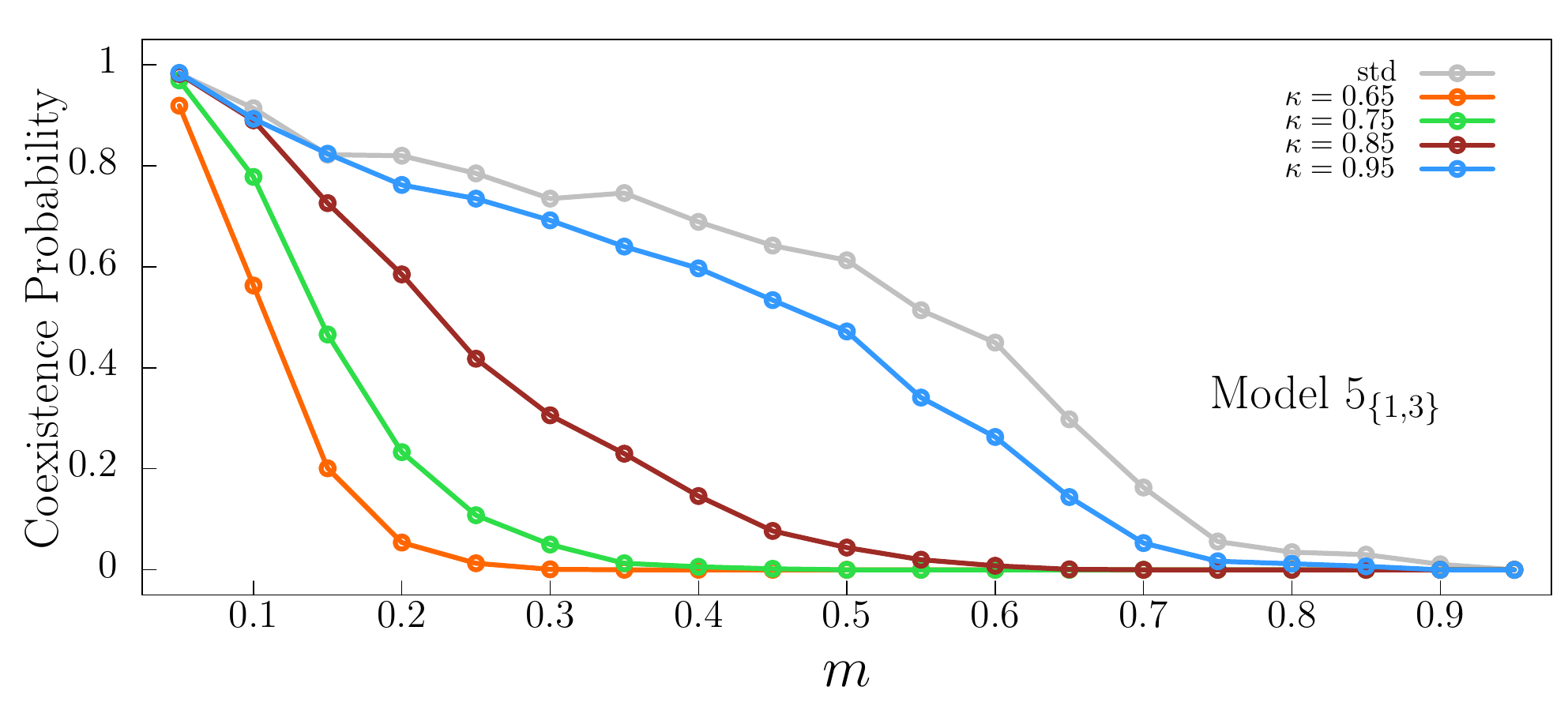}
        \caption{}\label{fig8b}
    \end{subfigure}
\caption{Coexistence probability as a function of the mobility $m$ for the generalised rock-paper-scissors with five species.
Figures \ref{fig8a}, and ~\ref{fig8b} depict the results for various $\kappa$, for models $5_{\{1,2\}}$ and $5_{\{1,3\}}$, respectively. In both figures, the orange, green, brown, and blue lines show the coexistence probability $\kappa=0.65$, $\kappa=0.75$, $\kappa=0.85$, $\kappa=0.95$, and $\kappa=1.0$, respectively; the grey line depicts the results for the standard model (without weak species).
The outcomes were calculated by sets of $1000$ simulations running for a timespan of $10000$ generations in lattices with $100^2$ grid sites, for $s=r=(1-m)/2$.}
  \label{fig8}
\end{figure}

\begin{figure}[t]
	\centering
	\includegraphics[width=45mm]{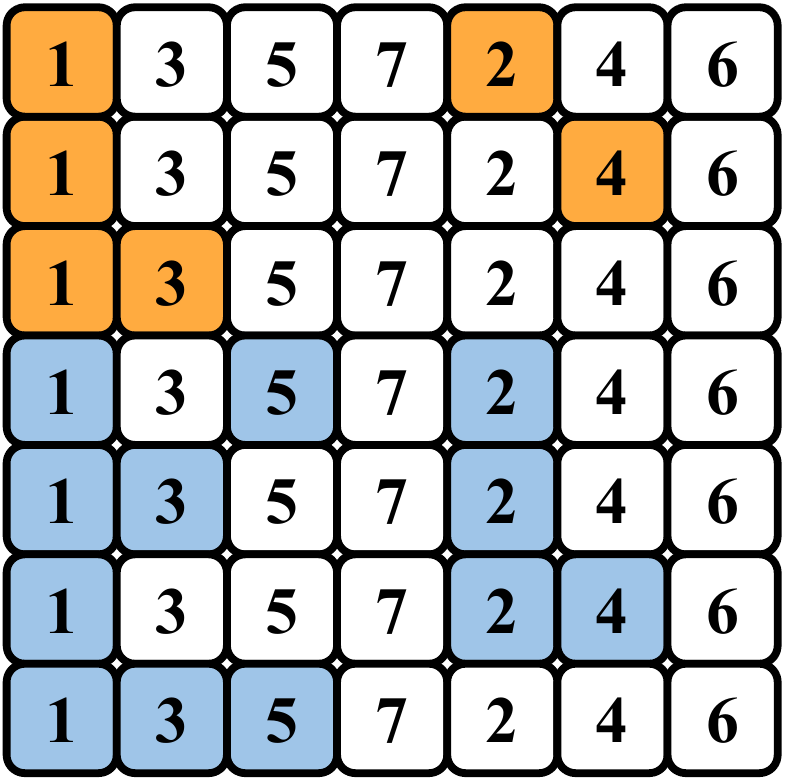}
    \caption{Illustration of the species' spatial displacement in the spatial patterns for the generalised rock-paper-scissors for $N=7$.
In the models for $n=2$, the weak species appearing in orange: 
$7_{\{1,2\}}$, $7_{\{1,3\}}$, and $7_{\{1,3\}}$. For $n=3$, the illustration shows the case $7_{\{1,2,5\}}$, $7_{\{1,2,3\}}$, $7_{\{1,2,4\}}$, and $7_{\{1,3,5\}}$, where the weak species shown in blue.}
  \label{fig9}
\end{figure}

\begin{figure}[h]
 \centering
        \begin{subfigure}{.48\textwidth}
        \centering
        \includegraphics[width=85mm]{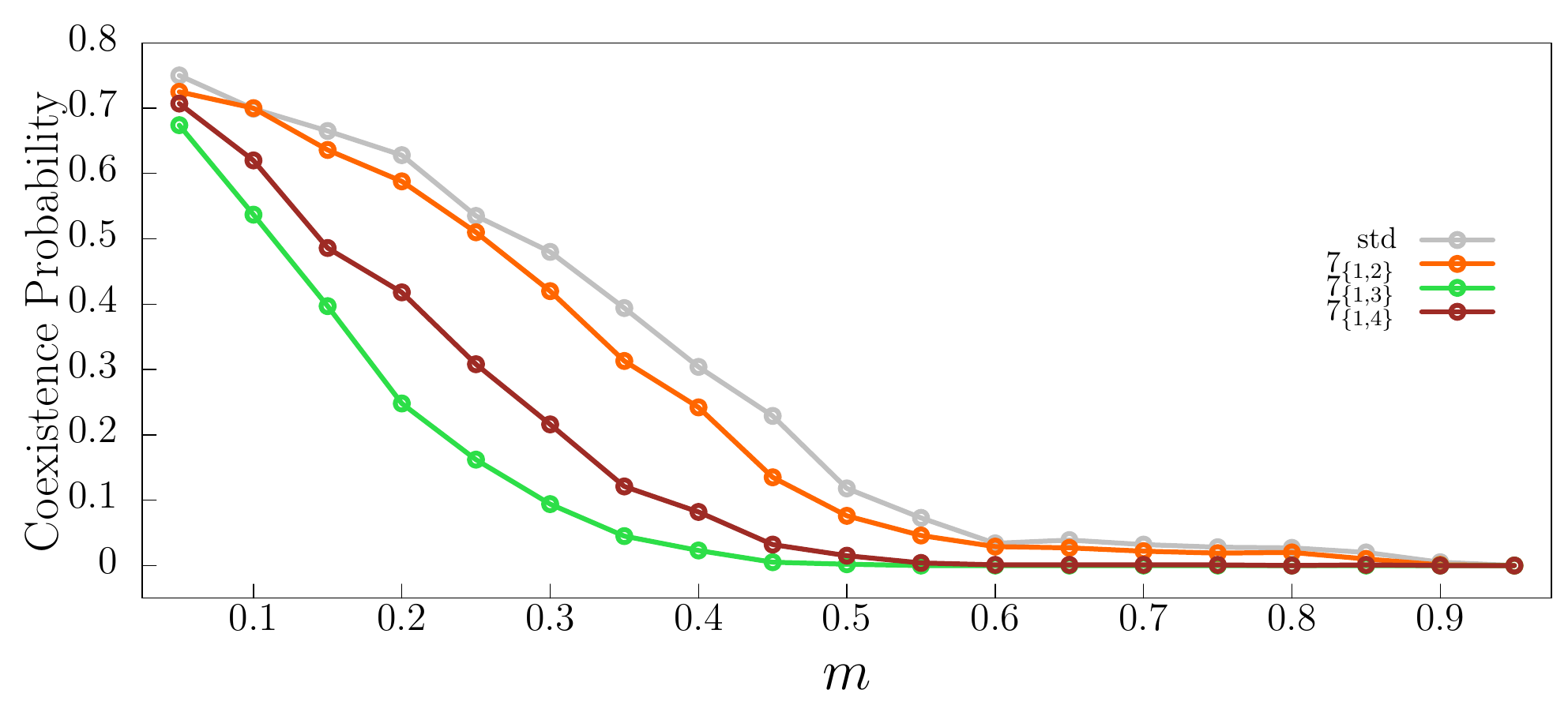}
        \caption{}\label{fig10a}
    \end{subfigure}\\
       \begin{subfigure}{.48\textwidth}
        \centering
        \includegraphics[width=85mm]{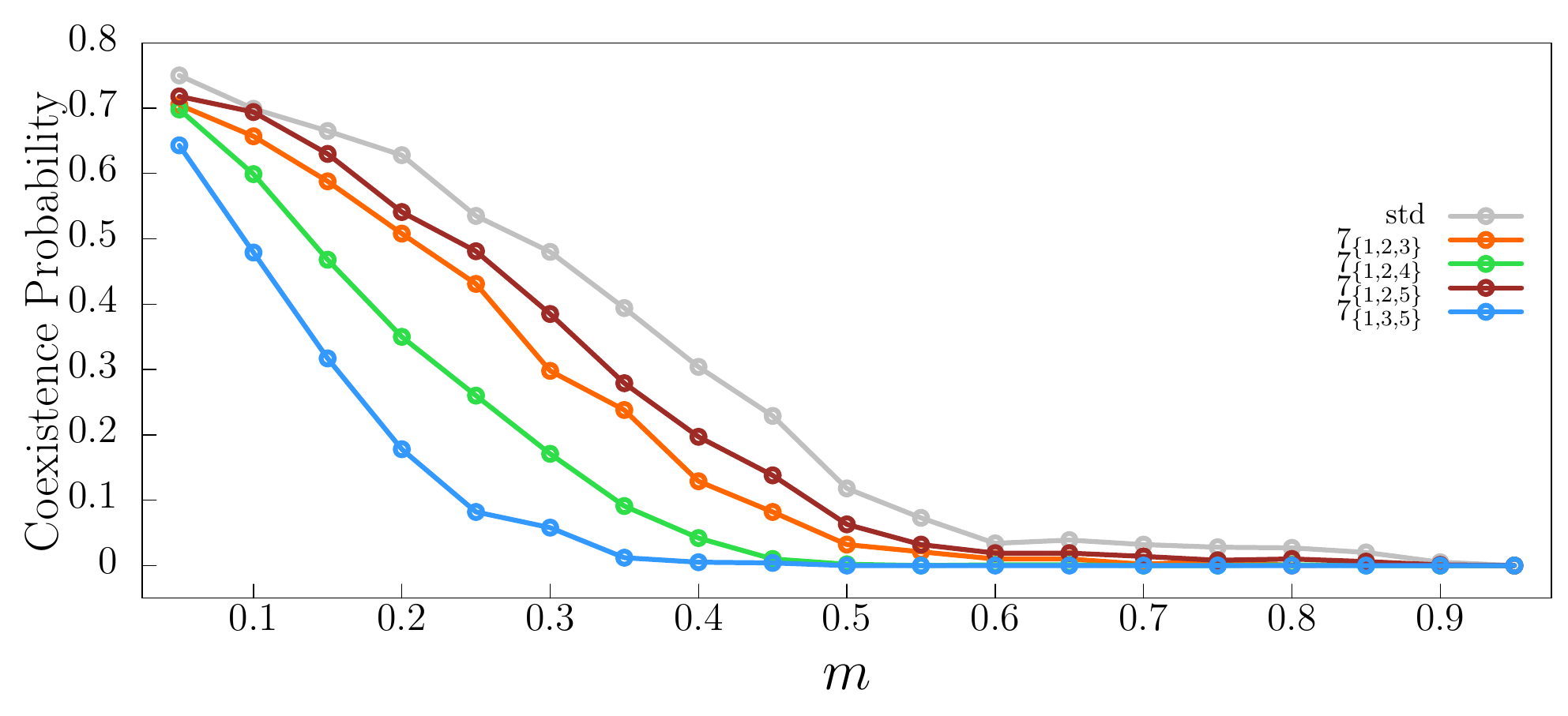}
        \caption{}\label{fig10b}
    \end{subfigure}
\caption{Coexistence probability as a function of the mobility $m$ for the generalised rock-paper-scissors with seven species.
Figure \ref{fig10a} and ~\ref{fig10b} show the outcomes for scenarios with two and three weak species, respectively. 
In Fig.~\ref{fig10a}, the orange, green, and brown lines depict the coexistence probability for models $7_{\{1,2\}}$, $7_{\{1,3\}}$, and $5_{\{1,4\}}$, respectively. The orange, green, brown, and blue lines in Fig.~\ref{fig10b} show the results for models $7_{\{1,2,3\}}$, $7_{\{1,2,4\}}$, $7_{\{1,2,5\}}$, and $7_{\{1,3,4\}}$, respectively. In both figures, the grey line shows the coexistence probability in the standard model, where all species are equally strong. The weak species' strength factor is $\kappa$; the interaction probabilities are $s=r=(1-m)/2$.
The results were obtained by running sets of $1000$ simulations until $t=10000$ generations in lattices with $100^2$ grid sites.}
  \label{fig10}
\end{figure}

\section{Autocorrelation function}
\label{sec4}

The spatial patterns presented in Figs.~\ref{fig3} and \ref{fig5} show concentrations of individuals of the same species asymmetrically distributed in the spiral arms. We now calculate
the scale of spatial domains occupied by each species using
the spatial autocorrelation function $C_i(r)$, with $i=1,2,3,4,5$, in terms of the radial coordinate $r$ \cite{Moura,PhysRevE.97.032415}.

We define the function $\phi_i(\vec{r})$ to describe the position $\vec{r}$ in the lattice occupied by individuals of species $i$. Calculating the mean value $\langle\phi_i\rangle$, we find the Fourier transform
\begin{equation}
\varphi_i(\vec{\kappa}) = \mathcal{F}\,\{\phi_i(\vec{r})-\langle\phi_i\rangle\},
\end{equation}
that gives the spectral densities
\begin{equation}
S_i(\vec{k}) = \sum_{k_x, k_y}\,\varphi_i(\vec{\kappa}).
\end{equation}

The autocorrelation function is found by employing the normalised inverse Fourier transform
\begin{equation}
C_i(\vec{r}') = \frac{\mathcal{F}^{-1}\{S_i(\vec{k})\}}{C(0)}.
\end{equation}
The spatial autocorrelation function for species $i$ as 
a function of the radial coordinate $r$ is then written as
\begin{equation}
C_i(r') = \sum_{|\vec{r}'|=x+y} \frac{C_i(\vec{r}')}{min\left[2N-(x+y+1), (x+y+1)\right]}.
\end{equation}
Finally, once the spatial autocorrelation is known, the
typical size of the spatial domains of organisms of species $i$ is calculated by 
assuming the threshold $C_i(l_i)=0.15$, where $l_i$ is the characteristic length scale for spatial domains of species $i$.

Figure \ref{fig5a} and \ref{fig5b} shows the autocorrelation function $C_i(r)$ for models $5_{\{1,2\}}$ and $5_{\{1,3\}}$, respectively. The outcomes were averaged from a set of $100$ simulations with different initial conditions, running in lattices with $500^2$ grid sites, running until $5000$ generations for $\kappa=0.75$; the error bars show the standard deviation and the colours follow the scheme in Fig.~\ref{fig1}. 
The dashed black line indicates the threshold for computing the characteristic length scale $l_i$ shown in the inset figures for several values of $\kappa$, namely, $0.75 \leq \kappa \leq 1.0$, with intervals $\Delta \kappa =0.05$. We consider the organisms' spatial organisation at $t=5000$ generations to compute the autocorrelation function; the interaction probabilities are $s=r=0.35$ and $m=0.3$.

The results show that the existence of multiple weak species influences the organisms' spatial organisation, with the asymmetry increasing as 
the strength factor $\kappa$ lowers.
According to Fig.~\ref{fig5a}, in model $5_{\{1,2\}}$, the 
regions dominated by species $2$ and $3$ are larger than the patches of other species. 
This happens because organisms of species $1$ and $2$ are weak; thus, individuals of $2$ and $4$ are killed at a lower rate than organisms of other species. 
Likewise, Fig.~\ref{fig5b} reveals that, due to the weakness of species $1$ and $3$, organisms of species $2$ and $4$ create the biggest groups in model $5_{\{1,3\}}$, with $l_2 > l_4$. 
\section{Species Densities}
\label{sec5}
Let us now investigate the impact of multiple weak species on population dynamics. First, we compute the temporal variation of the species densities in the simulations shown in Figs.~\ref{fig2} and ~\ref{fig4}. The densities of organisms of species $i$ are depicted in Figs.~\ref{fig6a} and ~\ref{fig6b}, for models $5_{\{1,2\}}$
and $5_{\{1,3\}}$, respectively.
The colours represent the species according to the scheme in Fig.~\ref{fig1}.

According to Fig.~\ref{fig6a}, the 
the pattern formation period lasts approximately $800$ generations if species $1$ and $2$ are weak.
After that, the average species densities are approximately constant, with fluctuations being inherent to the cyclic dominance of the rock-paper-scissors game. 
The outcomes show that as species $1$ is weak, organisms of species $2$ are less likely to be killed. This means that the population of species $2$ grows, predominating in the cyclic game.
However, despite the high number of enemies, species $3$ is not significantly affected by a population decrease because the organisms of species $2$ are weak. The consequence is that the density of species $3$ remains high, jeopardising the population of species $4$, which becomes the least abundant in the system. Finally, because of the low density of species $4$, organisms of species $5$ proliferate, being the second largest population, which affects species $1$, the second least populous.

The dynamics of species densities in the simulation of model $5_{\{1,3\}}$ show that the pattern formation stage lasts longer than in model $5_{\{1,2\}}$. According to the outcomes shown in Fig. \ref{fig6b}, initial fluctuations continue until approximately $t=1300$ generations. The results reveal that as organisms of species $1$ are weak, individuals of species $2$ are less likely to be eliminated, thus multiplying. The high density of species $2$ is harmful to species $3$ whose population declines. Besides being in smaller numbers, individuals of species $3$ are weak, which significantly benefits species $4$. Because of this, species $4$ is the most benefited, thus prevailing in the cyclic game. As a consequence, the number of organisms of species $5$ decreases because of the high number of opponents, benefiting species $1$: although conquering less territory due to the individuals' weakness, the population is preserved since the population of species $5$ is low. 

In summary, i) species $4$ predominates because the opponents are weak and less numerous; ii) species $2$ is the second more abundant since the adversary in the spatial game is weak; iii) species $1$ is in the third position because the population of adversary organisms is reduced; iv) species is the second least abundant since the opponents are more numerous; v) species $3$ is the least populous because organisms are in less number and weak.

\subsection{The role of weak species' strength in population dynamics} 

Our results conclude that species $2$ is the most benefitted if species $1$ and $2$ are weak, profiting with the reduced density of attacks suffered by individuals of species $1$. However, if species $1$ and $3$ are weak, the prevalence does not belong to any weak species but to species $3$.

We now explore the influence of the strength factor of the weak species in the population dynamics by calculating the average species densities $\rho_i$ for $0.7 \leq \kappa \leq 1.0$, with intervals $\Delta \kappa = 0.3$. We performed groups of $100$ realisations 
in grids of $500^2$ sites starting from different initial conditions, running until $5000$ generations.
Figures ~\ref{fig7a}, and ~\ref{fig7b} depict $\rho_i$ as functions of the time for models $5_{\{1,2\}}$ and $5_{\{1,3\}}$, respectively;
the colour follows the scheme in Fig.~\ref{fig1}.
The results were obtained by averaging the data from the second half of the simulations, thus, avoiding the initial fluctuation inherent to the spatial pattern formation process; the standard deviation is shown by error bars.

We found that the lower the
weak species' strength factor, the more unbalanced the spatial game. Furthermore, the results give evidence that model $5_{\{1,3\}}$ is more sensitive to the presence of weak species than model $5_{\{1,2\}}$. For example, for $\kappa=0.7$, the gap between the maximum and minimum average species densities is 
$\rho_4-\rho_5 \approx 0.11$ in model $5_{\{1,3\}}$, which is more than twice that in model $5_{\{1,2\}}$: $\rho_2-\rho_4 \approx 0.05$.
This happens because
the concentrations of species $1$ and $2$ are distant other. In contrast, patches with the largest number of individuals of species $1$ and $3$ are adjacent, thus maximising the effects of the unevenness in the cyclic game. 
\section{Coexistence Probability}
\label{sec6}
We now investigate the effects of multiple species in jeopardising biodiversity in the cyclic game models. Therefore, we implemented groups of 
$1000$ simulations in lattices with $100^2$ grid points for models $5_{\{1,2\}}$ and $5_{\{1,3\}}$, considering various 
values of $\kappa$. Each simulation started from different random initial conditions; coexistence occurs if at least one individual of all species is present at the end of the simulation. Therefore,
we define coexistence probability as the fraction of realisations resulting in coexistence - if at least one of the species is absent at the end of the simulation, extinction occurs. As we aim to quantify the coexistence probability in terms of the organisms' mobility, we repeated the simulations for $ 0.05\,<\,m\,<\,0.95$ in intervals of $ \Delta\, m\, =\,0.05$, with the selection and reproduction probabilities set to be $s\,=\,r\,=\,(1-m)/2$. 

Figures \ref{fig8a} and \ref{fig8b} shows the coexistence probability for models $5_{\{1,2\}}$ and $5_{\{1,3\}}$, respectively, with the grey line depicting the outcomes for the standard model (without weak species).
Orange, green, brown, and blue lines show the coexistence probability as a function of $m$ for $\kappa=0.65$, $\kappa=0.75$, $\kappa=0.85$, $\kappa=0.95$, and the standard model (where all organisms of every species are equally strong). In general, biodiversity is more threatened for systems with highly mobile individuals \cite{Reichenbach-N-448-1046,Moura}, even if there are no weak species (grey line). If multiple species are 
weak, the unevenness jeopardises biodiversity, reducing species' chances to coexist. Furthermore, our findings reveal that as species becomes weaker ($\kappa$ drops), biodiversity is more jeopardised in model $5_{\{1,3\}}$ than in model $5_{\{1,2\}}$, as shown, for example, by the green lines in Figs.~\ref{fig8a} and \ref{fig8b}. This is in agreement with the results shown in Figs.~\ref{fig7a} and \ref{fig7b}, which revealed that the reduction of the strength factor $\kappa$ leads to 
a more accentuated gap between the maximum and minimum species densities in model $5_{\{1,3\}}$ than in model $5_{\{1,2\}}$.

We conclude that the multiple weak species jeopardise more biodiversity if the organisms mostly occupy adjacent patches in the spiral patterns arising from random initial conditions. 
For $N=5$, the higher concentrations of individuals of the same species in the spiral arms are displaced in the following order: 
$\{(1, 3, 5, 2, 4)\}$. This means that: i) individuals of species $1$ and $2$ mostly live in patches that the most departed possible one of another, attenuating the impact over biodiversity in model $5_{\{1,2\}}$;
ii) organisms of species $1$ and $3$ mainly occupy adjacent spiral arms, which jeopardised biodiversity at a maximum level.

\subsection{Coexistence probability for the general case}

Our conclusions can be generalised for a system with an arbitrary odd number $N \geq 5$ of species with $n$ weak species, where $2 \leq n \leq (N-1)/2$. Overall,
the spatial patterns are spirals whose arms are composed of groups of individuals in the order $(i+2\,\beta)$, where $\beta$ is an integer that goes from $0$ to $N-1$ \cite{2012}. 
If $n$ weak multiple species are present, biodiversity is less jeopardised if they are departed from the most each other in the spatial patterns. 

As an illustration, Fig.~\ref{fig9} shows various scenarios for $N=7$ with two or three weak species. Accordingly, species occupy adjacent patches in the following order $\{1, 3, 5, 7, 2, 4, 6\}$.
For $n=2$, the weak species appear in orange in the cases
$7_{\{1,2\}}$, $7_{\{1,3\}}$, and $7_{\{1,3\}}$; for $n=3$, the illustration shows the case $7_{\{1,2,5\}}$, $7_{\{1,2,3\}}$, $7_{\{1,2,4\}}$, and $7_{\{1,3,5\}}$, where the weak species are highlighted in blue. 

To confirm our conclusions for the generalised rock-paper-scissors game with $7$ species, we performed sets of $1000$ simulations in grids with $100^2$ sites, considering the weak species' strength factor $\kappa=0.85$. The simulations ran until $10000$ generations; the fractions of realisations resulting in coexistence are shown in Figs.~\ref{fig10a} and ~\ref{fig10b} for $n=2$ and $n=3$, respectively. Mobility probability varies in the interval $0.05 \leq m \leq 0.95$ in intervals $\Delta m =0.05$.

First, for $n=2$, the weak species are further from each other in model $7_{\{1,2\}}$, thus jeopardising less biodiversity, as depicted by the orange line in Fig.~\ref{fig10a}. Now, suppose species $1$ and $4$ are weak. In that case, the shortest distance between organisms of both species is reduced, thus decreasing the chances for species to coexist, as verified by comparing the brown (model $7_{\{1,4\}}$) and orange lines (Model $7_{\{1,2\}}$) in Fig.~\ref{fig10a}. The worst scenario for biodiversity is if the high concentrations of both weak species are in adjacent patches, as occurs in model $7_{\{1,3\}}$. In this scenario, the coexistence probability reaches the minimum value, as depicted by the green line in Fig.~\ref{fig10a}.

Second, for $n=3$, there are more possibilities for disposing individuals of weak species in the spiral arm arising from the random initial conditions i) model $7_{\{1,2,5\}}$: all weak species are separated by individuals of other species, which represents the best scenario to biodiversity maintenance, as depicted by the brown line in Fig.~\ref{fig10b}; ii) model $7_{\{1,2,3\}}$: although individuals of species $2$ are far from species $1$ and $3$, patches dominated by species $1$ and $3$ are adjacent, thus decreasing the chance of species to coexist, as depicted by the orange line of Fig.~\ref{fig10b}; iii) model $7_{\{1,2,4\}}$: organisms of species $2$ and $4$ live in adjacent domains and are not far from the species $1$, which represents a more critical scenario for biodiversity maintenance, as depicted by the green line in Fig.~\ref{fig10b}; iv) model $7_{\{1,3,5\}}$: all three species occupy adjacent spiral arms, representing the scenario that biodiversity loss is more probable, as confirmed by the green line in Fig.~\ref{fig10b}.

Generally speaking, the outcomes confirm our prediction that: i) for $n=2$ the probability coexistence is maximum for model $7_{\{i,i+1\}}$, minimum for model $7_{\{i,i+2\}}$, and intermediate for model $7_{\{i,i+3\}}$; ii) for $n=3$, the case that jeopardises biodiversity the most is model $7_{\{i,i+2,i+4\}}$, while the minimum chance of biodiversity loss occurs in model $7_{\{i,i+1,i+4\}}$; model $7_{\{i,i+1,i+2\}}$ and model $7_{\{i,i+1,i+3\}}$
being intermediate, with the latter affecting more the coexistence probability, with $i=1,2,3,4,5$.
\section{Conclusions}
\label{sec7}

We investigated the generalised spatial rock-paper-scissors game with an arbitrary odd number $N$ of species whose organisms' spatial organisation arising from random initial conditions are spiral patterns. 
Among the species, $n$ are weak, with $2 \leq n \leq (N-1)/2$,
meaning that the organisms' selection capacity is lower than individuals of other species.
Initially, we ran stochastic simulations for $N=5$ and $n=2$ such that: i) species $i$ and $i+1$ are weak - organisms of species $i$ selecting individuals of species $i+1$; ii) species $i$ and $i+2$ are weak: - organisms of species $i$ neither do not kill nor is eliminated by individuals of species $i+2$. We conclude that the species whose individuals suffer less effective attacks have the chance to multiply and form the largest groups, thus occupying spatial domains with larger characteristic length scales. This yields a disequilibrium in territorial dominance, with the species abundance being determined by the position of the multiple weak species in the cyclic model.

We present a general prediction of the effects of multiple weak species in jeopardising biodiversity based on spatial patterns. For a system with $N$ species, organisms are distributed in spiral arms with high concentrations of species travelling in spiral arms, with waves of groups of species in the following order: $i$, $i+2$, ..., $i+N-1$, $i+1$, $i+3$, ..., $i-2$. Based on the species segregation, we conclude that if multiple weak species occupy adjacent spatial domains, the unevenness in the cyclic game is reinforced, maximising the chances of biodiversity loss. Therefore, 
the further apart the regions inhabited by different weak species are, the less the coexistence between the species is jeopardised.

Our findings may be useful for understanding general biological systems where various species are affected by external conditions. This may occur if an epidemic outbreak hits a system where not all species are immune to the disease-causing pathogen or when climate change alters environmental conditions, impacting the ability of various species to compete for natural resources.

 
\section*{Acknowledgments}
We thank CNPq, ECT, Fapern, and IBED for financial and technical support.

\bibliographystyle{elsarticle-num}
\bibliography{ref}

\begin{thebibliography}{10}
\expandafter\ifx\csname url\endcsname\relax
  \def\url#1{\texttt{#1}}\fi
\expandafter\ifx\csname urlprefix\endcsname\relax\def\urlprefix{URL }\fi
\expandafter\ifx\csname href\endcsname\relax
  \def\href#1#2{#2} \def\path#1{#1}\fi

\bibitem{ecology}
M.~Begon, C.~R. Townsend, J.~L. Harper, Ecology: from individuals to
  ecosystems, Blackwell Publishing, Oxford, 2006.

\bibitem{butterfly}
A.~Cormont, A.~H. Malinowska, O.~Kostenko, V.~Radchuk, L.~Hemerik, M.~F.
  WallisDeVries, J.~Verboom, Effect of local weather on butterfly flight
  behaviour, movement, and colonization: significance for dispersal under
  climate change, Biodiversity and Conservation 20 (2011) 483--503.

\bibitem{Nature-bio}
A.~Purvis, A.~Hector, Getting the measure of biodiversity, Nature 405 (2000)
  212--2019.

\bibitem{bacteria}
B.~C. Kirkup, M.~A. Riley, Antibiotic-mediated antagonism leads to a bacterial
  game of rock-paper-scissors in vivo, Nature 428 (2004) 412--414.

\bibitem{Coli}
B.~Kerr, M.~A. Riley, M.~W. Feldman, B.~J.~M. Bohannan, Local dispersal
  promotes biodiversity in a real-life game of rock–paper–scissors, Nature
  418 (2002) 171.

\bibitem{Allelopathy}
R.~Durret, S.~Levin, Allelopathy in spatially distributed populations, J.
  Theor. Biol. 185 (1997) 165--171.

\bibitem{lizards}
B.~Sinervo, C.~M. Lively, The rock-scissors-paper game and the evolution of
  alternative male strategies, Nature 380 (1996) 240--243.

\bibitem{Extra1}
I.~Volkov, J.~R. Banavar, S.~P. Hubbell, A.~Maritan, Patterns of relative
  species abundance in rainforests and coral reefs, Nature 450 (2007) 45.

\bibitem{doi:10.1098/rsif.2014.0735}
A.~Szolnoki, M.~Mobilia, L.-L. Jiang, B.~Szczesny, A.~M. Rucklidge, M.~Perc,
  Cyclic dominance in evolutionary games: a review, Journal of The Royal
  Society Interface 11~(100) (2014) 20140735.

\bibitem{PARK2023113004}
J.~Park, X.~Chen, A.~Szolnoki, Competition of alliances in a cyclically
  dominant eight-species population, Chaos, Solitons \& Fractals 166 (2023)
  113004.

\bibitem{PhysRevE.93.062307}
A.~Szolnoki, M.~c.~v. Perc, Zealots tame oscillations in the spatial
  rock-paper-scissors game, Phys. Rev. E 93 (2016) 062307.

\bibitem{doi:10.1063/5.0093342}
Y.~Lu, C.~Shen, M.~Wu, C.~Du, L.~Shi, J.~Park, Enhancing coexistence of mobile
  species in the cyclic competition system by wildlife refuge, Chaos: An
  Interdisciplinary Journal of Nonlinear Science 32~(8) (2022) 081104.

\bibitem{KABIR2021125767}
K.~A. Kabir, J.~Tanimoto, The role of pairwise nonlinear evolutionary dynamics
  in the rock–paper–scissors game with noise, Applied Mathematics and
  Computation 394 (2021) 125767.

\bibitem{doi:10.1063/5.0102416}
J.~Park, Correlation between the formation of new competing group and spatial
  scale for biodiversity in the evolutionary dynamics of cyclic competition,
  Chaos: An Interdisciplinary Journal of Nonlinear Science 32~(8) (2022)
  081101.

\bibitem{Reichenbach-N-448-1046}
T.~Reichenbach, M.~Mobilia, E.~Frey, Mobility promotes and jeopardizes
  biodiversity in rock-paper-scissors games, Nature 448 (2007) 1046--1049.

\bibitem{Rev1}
A.~Szolnoki, M.~Mobilia, L.-L. Jiang, B.~Szczesny, A.~M. Rucklidge, M.~Perc,
  Cyclic dominance in evolutionary games: a review, Journal of The Royal
  Society Interface 11~(100) (2014) 20140735.

\bibitem{Bazeia_2017}
D.~Bazeia, J.~Menezes, B.~F. de~Oliveira, J.~G. G.~S. Ramos, Hamming distance
  and mobility behavior in generalized rock-paper-scissors models, Europhysics
  Letters 119~(5) (2017) 58003.

\bibitem{Avelino-PRE-89-042710}
P.~P. Avelino, D.~Bazeia, L.~Losano, J.~Menezes, B.~F. de~Oliveira, Interfaces
  with internal structures in generalized rock-paper-scissors models, Phys.
  Rev. E 89 (2014) 042710.

\bibitem{Menezes_2022}
J.~Menezes, B.~Ferreira, E.~Rangel, B.~Moura, Adaptive altruistic strategy in
  cyclic models during an epidemic, Europhysics Letters 140~(5) (2022) 57001.

\bibitem{Rev6}
A.~Szolnoki, J.~Vukov, M.~c.~v. Perc, From pairwise to group interactions in
  games of cyclic dominance, Phys. Rev. E 89 (2014) 062125.

\bibitem{PhysRevE.99.052310}
P.~P. Avelino, J.~Menezes, B.~F. de~Oliveira, T.~A. Pereira, Expanding spatial
  domains and transient scaling regimes in populations with local cyclic
  competition, Phys. Rev. E 99 (2019) 052310.

\bibitem{Rev4}
A.~Szolnoki, M.~Perc, Vortices determine the dynamics of biodiversity in
  cyclical interactions with protection spillovers, New Journal of Physics
  17~(11) (2015) 113033.

\bibitem{PARKCHAOS}
J.~Park, Y.~Do, B.~Jang, Multistability in the cyclic competition system, Chaos
  28 (2018) 113110.

\bibitem{Nagatani2018}
H.~Cheng, N.~Yao, Z.-G. Huang, J.~Park, Y.~Do, Y.-C. Lai, Heterogeneous network
  promotes species coexistence: metapopulation model for rock-paper-scissors
  game, Scientific Reports 8 (2018) 2045--2322.

\bibitem{Park_2019}
J.~Park, Fitness-based mutation in the spatial rock-paper-scissors game:
  Shifting of critical mobility for extinction, {EPL} (Europhysics Letters)
  126~(3) (2019) 38004.

\bibitem{RANGEL2022104689}
Combination of survival movement strategies in cyclic game systems during an
  epidemic, Biosystems 217 (2022) 104689.

\bibitem{PhysRevE.105.024309}
P.~P. Avelino, B.~F. de~Oliveira, R.~S. Trintin, Lotka-volterra versus
  may-leonard formulations of the spatial stochastic rock-paper-scissors model:
  The missing link, Phys. Rev. E 105 (2022) 024309.

\bibitem{Avelino_2018}
P.~P. Avelino, D.~Bazeia, L.~Losano, J.~Menezes, B.~F. de~Oliveira, Spatial
  patterns and biodiversity in off-lattice simulations of a cyclic
  three-species lotka-volterra model, {EPL} (Europhysics Letters) 121~(4)
  (2018) 48003.

\bibitem{Avelino-PRE-86-031119}
P.~P. Avelino, D.~Bazeia, L.~Losano, J.~Menezes, von neummann's and related
  scaling laws in rock-paper-scissors-type games, Phys. Rev. E 86 (2012)
  031119.

\bibitem{2012}
P.~P. Avelino, D.~Bazeia, L.~Losano, J.~Menezes, B.~F. Oliveira, Junctions and
  spiral patterns in generalized rock-paper-scissors models, Phys. Rev. E 86
  (2012) 036112.

\bibitem{Park2017}
J.~Park, Y.~Do, B.~Jang, Y.-C. Lai, Emergence of unusual coexistence states in
  cyclic game systems, Scientific Reports 7~(1) (2017) 2045--2322.

\bibitem{Pereira}
T.~A. Pereira, J.~Menezes, L.~Losano, Interface networks in models of competing
  species, Intern. J. of Mod., Sim. and Sci. Comp. 9 (2018) 1850046.

\bibitem{Menezes_2022A}
J.~Menezes, B.~Moura, E.~Rangel, Adaptive survival movement strategy to local
  epidemic outbreaks in cyclic models, Journal of Physics: Complexity 3~(4)
  (2022) 045008.

\bibitem{MENEZES2022104777}
Spatial organisation plasticity reduces disease infection risk in
  rock–paper–scissors models, Biosystems 221 (2022) 104777.

\bibitem{TENORIO2022112430}
M.~Tenorio, E.~Rangel, J.~Menezes, Adaptive movement strategy in
  rock-paper-scissors models, Chaos, Solitons \& Fractals 162 (2022) 112430.

\bibitem{Moura}
B.~Moura, J.~Menezes, Behavioural movement strategies in cyclic models,
  Scientific Reports 11 (2021) 6413.

\bibitem{Anti1}
J.~Menezes, Antipredator behavior in the rock-paper-scissors model, Phys. Rev.
  E 103 (2021) 052216.

\bibitem{Anti2}
J.~Menezes, B.~Moura, Mobility-limiting antipredator response in the
  rock-paper-scissors model, Phys. Rev. E 104 (2021) 054201.

\bibitem{MENEZES2022101606}
J.~Menezes, E.~Rangel, B.~Moura, Aggregation as an antipredator strategy in the
  rock-paper-scissors model, Ecological Informatics 69 (2022) 101606.

\bibitem{Rev2}
A.~Szolnoki, M.~c.~v. Perc, G.~Szab\'o, Defense mechanisms of empathetic
  players in the spatial ultimatum game, Phys. Rev. Lett. 109 (2012) 078701.

\bibitem{Rev3}
A.~Szolnoki, M.~c.~v. Perc, Correlation of positive and negative reciprocity
  fails to confer an evolutionary advantage: Phase transitions to elementary
  strategies, Phys. Rev. X 3 (2013) 041021.

\bibitem{uneven}
J.~Menezes, B.~Moura, T.~A. Pereira, Uneven rock-paper-scissors models:
  Patterns and coexistence, Europhysics Letters 126~(1) (2019) 18003.

\bibitem{weakest}
M.~Frean, E.~R. Abraham, Rock–scissors–paper and the survival of the
  weakest, Proc. R. Soc. Lond. B. 268 (2001) 1323--1327.

\bibitem{PedroWeak}
P.~P. Avelino, B.~F. de~Oliveira, R.~S. Trintin, Predominance of the weakest
  species in lotka-volterra and may-leonard formulations of the
  rock-paper-scissors model, Phys. Rev. E 100 (2019) 042209.

\bibitem{Weak4}
P.~P. Avelino, B.~F. de~Oliveira, R.~S. Trintin, Performance of weak species in
  the simplest generalization of the rock-paper-scissors model to four species,
  Phys. Rev. E 101 (2020) 062312.

\bibitem{parity}
P.~Avelino, B.~{de Oliveira}, R.~Trintin, Parity effects in rock-paper-scissors
  type models with a number of species $ns \leq12$, Chaos, Solitons \& Fractals
  155 (2022) 111738.

\bibitem{doi:10.1063/5.0106165}
J.~Menezes, S.~Batista, M.~Tenorio, E.~Triaca, B.~Moura, How local antipredator
  response unbalances the rock-paper-scissors model, Chaos: An
  Interdisciplinary Journal of Nonlinear Science 32~(12) (2022) 123142.

\bibitem{leonard}
R.~M. May, W.~J. Leonard, Nonlinear aspects of competition between three
  species, SIAM J. Appl. Math. 29 (1975) 243--253.

\bibitem{PhysRevE.97.032415}
P.~P. Avelino, D.~Bazeia, L.~Losano, J.~Menezes, B.~F. de~Oliveira, M.~A.
  Santos, How directional mobility affects coexistence in rock-paper-scissors
  models, Phys. Rev. E 97 (2018) 032415.

\end{thebibliography}

\end{document}